\documentclass[%
 reprint,
 superscriptaddress,
 amsmath,amssymb,
 aps,
]{revtex4-2}

\usepackage{graphicx}
\usepackage{dcolumn}
\usepackage{bm}
\usepackage{hyperref}
\usepackage[caption=false]{subfig}
\usepackage{makecell}
\usepackage[export]{adjustbox}
\usepackage{multirow}

\begin{document}

\preprint{APS/123-QED}

\title{Nagaoka Ferromagnetism in $3\times 3$ Arrays and Beyond}
\author{Yan Li}
\email{yan.li@nist.gov}
\affiliation{Joint Quantum Institute, \\
University of Maryland and National Institute of Standards and Technology, College Park, Maryland 20742, USA}

\author{Keyi Liu}%
\affiliation{Joint Quantum Institute, \\
University of Maryland and National Institute of Standards and Technology, College Park, Maryland 20742, USA}
\author{Garnett W. Bryant}
\affiliation{Joint Quantum Institute, \\
University of Maryland and National Institute of Standards and Technology, College Park, Maryland 20742, USA}
\affiliation{Nanoscale Device Characterization Division,\\
National Institute of Standards and Technology, Gaithersburg, Maryland 20899, USA}

\date{\today}

\begin{abstract}
Nagaoka ferromagnetism (NF) is a long-predicted example of itinerant ferromagnetism (IF) in the Hubbard model that has been studied theoretically for many years. The condition for NF, an infinite on-site Coulomb repulsion and a single hole in a half-filled band, does not arise naturally in materials. NF was only realized recently for the first time in experiments on a $2\times 2$ array of gated quantum dots. Dopant arrays and gated quantum dots in Si allow for engineering controllable systems with complex geometries. This makes dopant and quantum dot arrays good candidates to study NF in different array geometries through analog quantum simulation. Here we present theoretical simulations done for $3\times 3$ arrays and larger $N\times N$ arrays, and predict the emergence of different forms of ferromagnetism in different geometries. We find NF in perfect $3\times 3$ arrays, as well as in $N\times N$ arrays for one hole doping of a half-filled band. The ratio of the  hopping $t$ to Hubbard on-site repulsion $U$ that defines the onset of NF scales as $1/N^{4}$ as $N$ increases, approaching the bulk limit of infinite $U$ for large $N$. Additional simulations are done for geometries made by removing sites from $N\times N$ arrays. Different forms of ferromagnetism are found for different geometries. Loops show ferromagnetism, but only for three electrons. For loops, the critical $t/U$ for the onset of ferromagnetism scales as $N$ as the loop length increases. We show that the different dependences on size for loops and $N\times N$ arrays can be understood by scaling arguments that highlight the different energy contributions to each different form of ferromagnetism. Our results show how analog quantum simulation with small arrays can elucidate the role of effects including wavefunction connectivity; system geometry, size and symmetry; bulk and edge sites; and kinetic energy in determining quantum magnetism of small systems. 
\end{abstract}
\maketitle

\section{\label{intro} Introduction}
Nagaoka ferromagnetism (NF) is a well-known result of the Hubbard model, where Nagaoka \cite{Nagaoka_original} proved rigorously the existence and uniqueness of a saturated, ferromagnetic ground state in a single-band Hubbard model on a lattice. In the case of one hole in a half-filled band and infinite, repulsive, same-site Coulomb interaction $U$ between opposite spins, the ground state has the maximum total spin $S$. This comes from a non-trivial interplay between quantum dynamics and Coulomb interaction, where the kinetic and repulsive energy terms compete resulting in a ferromagnetic ground state. This competition is determined by how the single hole can move around the lattice.  

NF is one of the exact results for itinerant ferromagnetism (IF) and has been studied theoretically for years in various conditions \cite{tasaki1998,lieb1989}. However, the specific conditions proved by Nagaoka do not arise in natural materials and was observed experimentally for the first time in an engineered $2\times 2$ quantum dot plaquette \cite{dehollain2020}. It remains an open question when NF can be extended to larger, finite-systems that have internal as well as edge sites, how NF would behave in realistic conditions where $U$ is large but not infinite, and how NF is affected by long-range interactions or disorder. 

Following Nagaoka's original work, many have tried to generalize Nagaoka’s result. Tasaki provided a proof of a generalized version of Nagaoka's theorem \cite{tasaki1998}. Mielke and others found a new class of ferromagnetic states in the Hubbard model which they called "flat-band ferromagnetism" \cite{mielke1999}. The possibility of extending NF to finite $U$ with finite densities of holes has also been studied for different infinite lattices \cite{Hanisch1995, barbieri1990}, for large 2D lattices \cite{becca2001}, and for square lattices \cite{liu2012}. also see \cite{Maifm2012} for Nagaoka polaron regime for finite hole density. There have also been many theoretical works done on the instability of NF, that show when NF does not take place \cite{suto1991, toth1991, doucot1989}. Additional theoretical work has elucidated the polaronic correlation around holes in NF \cite{Maifm2012, samajdar2023, samajdar2024}. 

In recent years, as computational techniques have advanced, there has been more discussion of small, finite systems that exhibit NF. For example, finite lattices with up to 24 sites with periodic boundary conditions were studied using quantum Monte Carlo \cite{QMC1,QMC2}. Different geometries of finite systems were studied using exact diagonalization techniques (ED) \cite{Buterakos2019, Buterakos2023}. These studies revealed the importance of geometry for realizing NF in finite systems, which could be achievable with current experiments using gated quantum dot or dopant arrays.

Quantum dots, also known as artificial atoms, can be used to perform simulations beyond what classical computers can do \cite{hensgens2017,barthelemy2013,salfi2016,cirac2012}. Among various experimental platforms, dopant-based quantum dots have many advantages, such as the precise placement of dopant atoms and tunable site-to-site hopping, making them a strong candidate for simulating strongly correlated systems, like the Hubbard model in the large $U$ limit. Silver recently fabricated $3\times 3$ arrays of single/few-dopant quantum dots and demonstrated analog quantum simulation of a 2D Fermi-Hubbard model with tunable parameters \cite{wang2022}. This was the first two-dimensional quantum-dot simulation with internal sites. This provides an experimental pathway to probe NF in $3\times 3$ arrays and in many other finite geometries. Gated semiconductor quantum dots show similar promise. The original effort to study NF in a $2\times 2$ plaquette \cite{dehollain2020} has been extended to study exciton and spin dynamics in $2\times 4$ ladders. \cite{hsiao2024, Zhang2023} Linear chains formed from multidopant quantum dot clusters has been used to study topological states in the Su-Schrieffer-Heeger model. \cite{kiczynski2022}

In this paper, we simulate $3\times 3$ dopant arrays in the conditions required for NF, i.e. one less electron than a half-filled band and in the large $U$ limit. We predict that NF will exist in a $3\times 3$ array, as in a $2\times 2$ array, as well as in $N\times N$ arrays, for a finite region of $t/U$, where $t$ is the nearest-neighbor hopping. For $t/U$ below the transition point, the ground state of the system has maximum total spin $S$, i.e. the ground state is ferromagnetic. Since the direction of the ferromagnetism is arbitrary, the system energy only depends on total spin $S$ but not the spin in the $z$-direction ($s_z$). The ferromagnetic state with maximum $s_z = S$ with all of the spins aligned in the same $z$ direction is degenerate with all other ferromagnetic states with smaller $s_z$ as long as they have the same total $S$. The ground state is $(2S+1)$-fold degenerate when there is Nagaoka ferromagnetism. Above the transition point, the ground state is no longer the maximum-total-spin state and is no longer ferromagnetic. We also simulate variations of the perfect $3\times 3$ array to determine when NF is robust to experimental impurities or disorder.

The quantum nature of magnetism plays an essential role in determining how and when ferromagnetism occurs. The small-scale simulators that we consider in this paper provide an excellent testbed for investigating quantum effects. We find that the connectivity of the many-body wavefunction plays a key role in determining when NF can occur. Wavefunction connectivity also determines when forms of itinerant ferromagnetism different from NF at fillings other than one hole in a half-filled band can occur. Band filling is another key condition that determines the type of ferromagnetism. We will show that very different fillings, related to the number of holes in a half-filled band or the number of electrons in an empty band, will define the possibility of ferromagnetism in similar, related structures. The competition between kinetic energy and coulomb repulsion effects determines where in $t/U$ space the transition to ferromagnetism occurs. The $t/U$ ratio at the transition can increase or decrease with system size, depending on the array geometry (and the related wavefunction connectivity), and the band filling. We develop simple scaling arguments to explain both types of size-dependence and the critical role that the kinetic energy plays, through its dependence on band filling, in determining the form of the scaling. The role of these quantum effects will be highlighted as we discuss NF in different small arrays to show how quantum simulation with small arrays could reveal these critical effects. It should be emphasized that dopant arrays, once they can be made perfectly \cite{Wyrick} would be a versatile solid-state platform to explore the diversity of arrays and the range of effects that we discuss. Gated quantum dots will also be a versatile solid-state platform as techniques to control dots in complex geometries are developed.    

The paper is structured as follows: In Section \ref{2dhubbard}, we discuss the 2D Hubbard model, the signatures of finding NF in finite systems, and how the systems are modelled in our simulations. In Section \ref{Sq_arr}, we present our simulation results for the $3\times 3$ and $N\times N$ square arrays, which exhibit NF up to a finite $t/U$ transition point that decreases with increasing $N$. In Section \ref{robustness}, we describe the robustness of NF in a $3\times 3$ array with disordered sites, that is sites that have been removed. Most importantly, we show that a loop array, i.e. an $N\times N$ with the internal sites removed, possesses a ferromagnetic ground state for three-electron filling with a $t/U$ transition that scales linearly with loop length $N$, in stark contrast to NF in perfect, square arrays. In this section, we also present results to show how the $t/U$ ratio at the transition to ferromagnetism scales with system size $N$.  Scaling arguments explain this dependence on $N$ and elucidate the competition between kinetic energy and repulsive interaction affects to determines when the transition occurs. Finally in Section \ref{summary}, we summarize and discuss essential results about quantum magnetism that can be revealed by simulations done on small arrays of dopants or quantum dots.

\section{\label{2dhubbard} Hubbard Model}
NF occurs for electrons described by the Hubbard model, which takes the form as follows 

\begin{equation}
\mathcal{H} = t\sum_{<i,j>, \sigma} c^{\dag}_{i,\sigma}c_{j,\sigma} + U\sum_i n_{i,\uparrow} n_{i,\downarrow}
\end{equation}
 where $t$ is the hopping between nearest neighbor sites, $c^{\dag}$ and $c$ are the creation and annihilation operators, $n$ is the particle number operator, and $U$ is the on-site interaction between spin-up and spin-down electrons. NF is determined by a competition between hopping energy and repulsion, hence the parameter that matters is $t/U$. Some use the hopping term $t$ with a minus sign, but here we use the convention without a minus sign and allow the value of $t$ to be positive or negative. The sign of $t$ is important because in some systems the Nagaoka theorem can be proved for one sign but not the other. The sign of t can often be changed with a basis change. In particular, the transformation between the signs of $t$ can be done for a bipartite system. In these cases, NF, if it occurs, will occur for both signs of $t$. The $N\times N$ arrays we study are bipartite lattices.

\begin{figure}[h]
\subfloat[$S$ states]{%
  \includegraphics[scale=0.045,left]{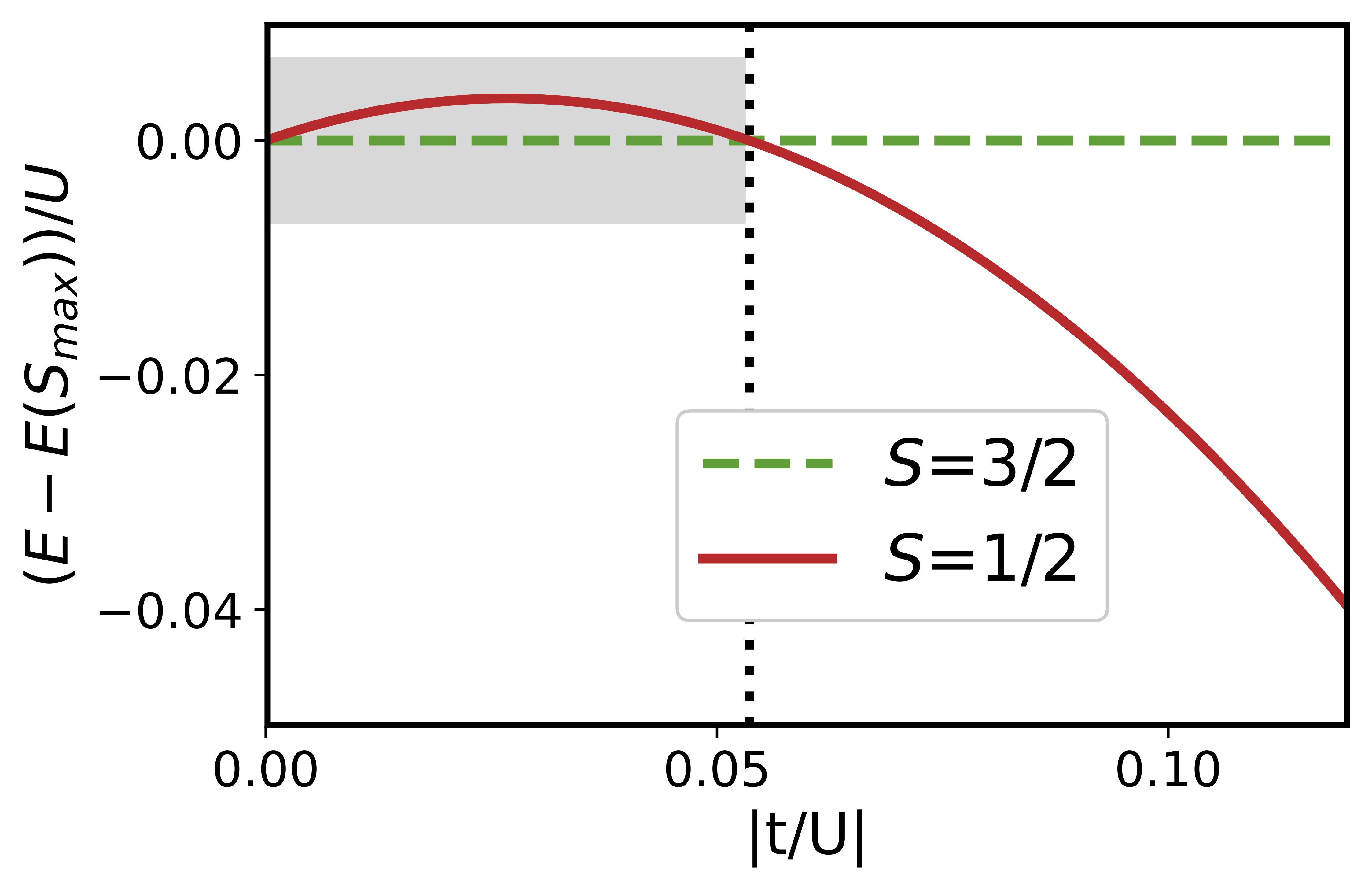}%
  \label{fig:2x2_totalS}
}\\
\subfloat[$s_z$ states]{%
  \includegraphics[scale=0.045,left]{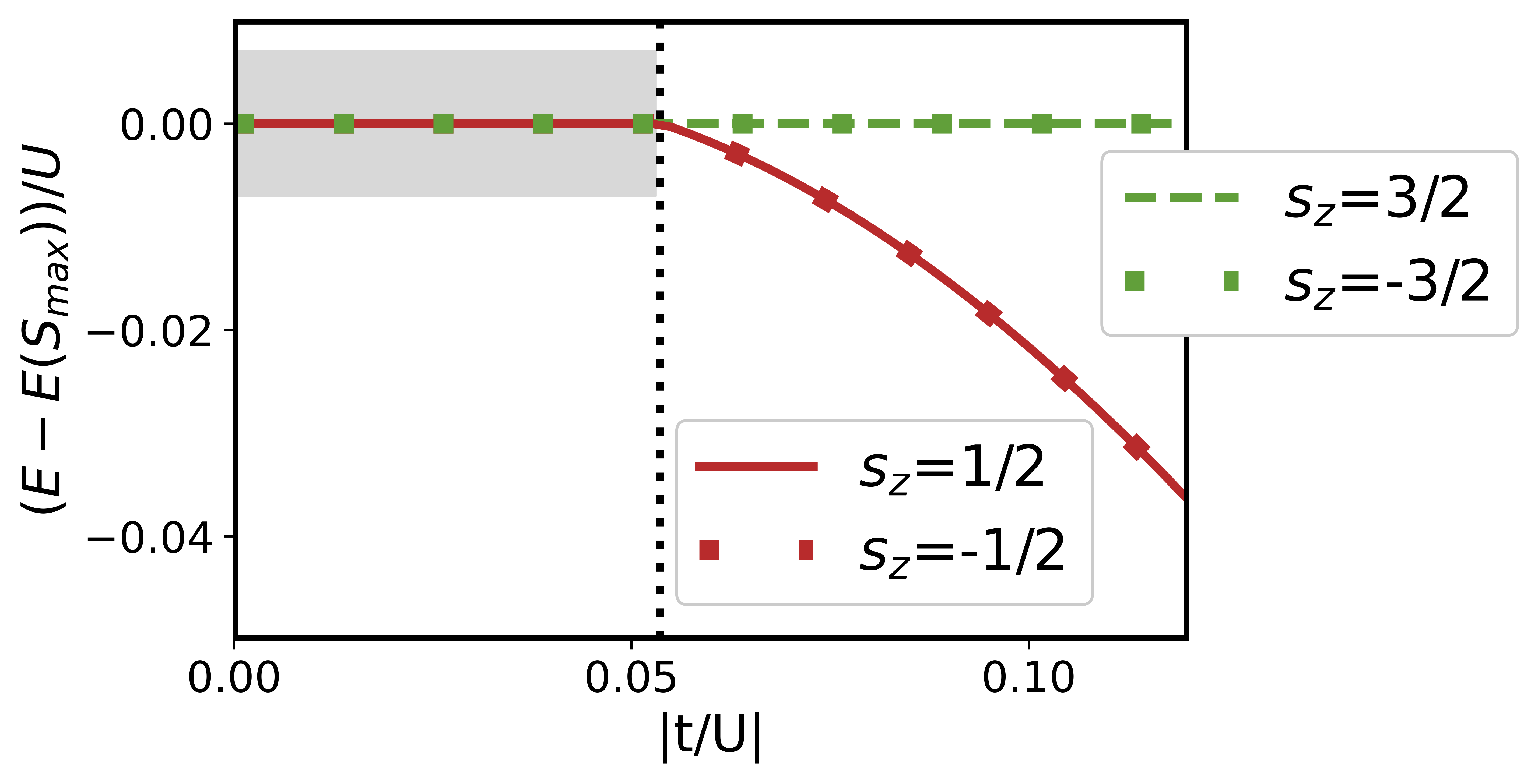}%
  \label{fig:2x2_sz}
}
\caption{Nagaoka ferromagnetism in a $2\times 2$ array. (a) Energy spectrum ($S=3/2$ and $S=1/2$ as a function of $|t/U|$), which are offset by the energy of $S=3/2$. Black dotted line indicates the transition for NF. (b) Ground state for each of the different spin z subspaces as a function of $|t/U|$, which are offset by the energy for $s_z = 3/2$. Above and below the transition point, the ground state for $s_z=\pm 1/2$ switches from total $S=3/2$ to $S=1/2$. NF is shown by the shaded region.}
\label{fig:2x2_en}
\end{figure}

To find out if a system has NF, it is necessary to determine whether the one-hole ground state has maximum total spin. For a finite system like the $2\times 2$ array with 3 electrons, the Hamiltonian is small and the energy spectrum can be calculated directly, as shown in \cite{dehollain2020}. The energies for the two different total spins, $S = 3/2$ and $S = 1/2$ are plotted against the parameter $t/U$ in Fig. \ref{fig:2x2_totalS}. Both energies are shifted with respect to the energy of the maximum-total-spin state $E_{S=3/2}= -2t$. Below $|t/U| \approx 0.053$, the ground state of the system is the maximum-total-spin $S_{max} = 3/2$ state. In the ferromagnetic ground state, the states with $S=S_{max}$ have a degeneracy of $2S_{max} + 1$. In the $2\times 2$ array with 3 electrons, there are four degenerate $s_z$ states with $s_z = -3/2, -1/2, 1/2, 3/2$. Because the Hubbard model does not include spin-mixing, ED can be done for each $s_z$ subspace, thereby reducing the size of each diagonalization. In our simulation, instead of plotting the ground state energy for each total spin $S$, we plot the ground state energy for the different spin-$z$ subspaces as shown in Fig. \ref{fig:2x2_sz}, and shift them by the energy of the state with maximum $s_z$ $E_{s_z=3/2} = -2t$. Below the transition when the system is in the NF regime, the ground state is $4$-fold degenerate. The signature of NF is when the ground states of the spin-$z$ subspaces become degenerate with each other. This guarantees that the ground state of the system has maximum total spin. 

Fig. \ref{fig:2x2_en} shows how energy behaves as $|t/U|$ varies. The lattice is bipartite so the sign of $t$ does not matter here, and $t$ and $-t$ give the same energies. This is not always the case as seen later when the system is no longer bipartite, and the $\pm t$ cases can behave very differently. In Fig.~\ref{fig:2x2_sz}, the $s_z = \pm  1/2$ state is always a ground state, but the ground state after the transition has a different total $S$ and degeneracy compared to before the transition. This change in total spin can be verified by direct calculation of the total spin for each spin $z$. 

The existence of NF in $2\times 2$ array has been verified experimentally in \cite{dehollain2020} for a $2\times 2$ gated quantum dot plaquette. An ab initio calculation of NF in quantum dots was done in \cite{wang2019}. Its robustness was discussed in \cite{wang2019} and theoretically in \cite{Buterakos2019}.  

\section{\label{Sq_arr} Larger Square Arrays}
\subsection{\label{sec:3x3} $3\times 3$ case}

The $2\times 2$ array has no internal sites. However, the $3\times 3$ dopant array studied by \cite{wang2022} has one internal site. In this section, we look at $3\times 3$ and larger arrays to see how NF behaves as the size of the system increases and the system approaches the 2D bulk limit. The largest arrays consider here are $8\times 8$ arrays, which should be accessible with dopant arrays. These arrays have more internal sites than edge sites and should represent the beginning of transition to the 2D bulk limit. Previous work \cite{white2001} has been done on $9\times 9$ systems for a t-J model in the polaron picture.

Our results will define the $t/U$ ratios needed for NF. For the $3\times 3$ array, the ground state energy for each spin-$z$ subspace is calculated to determine the presence of NF, and if it exists, to identify the point at which the transition occurs. The half-filled band with one hole has $8$ electrons, and the maximum total spin is $4$. If all $9 s_z$ subspaces have the same ground-state energy, the system has NF. 

\begin{figure}[h]
    \includegraphics[scale=0.05]{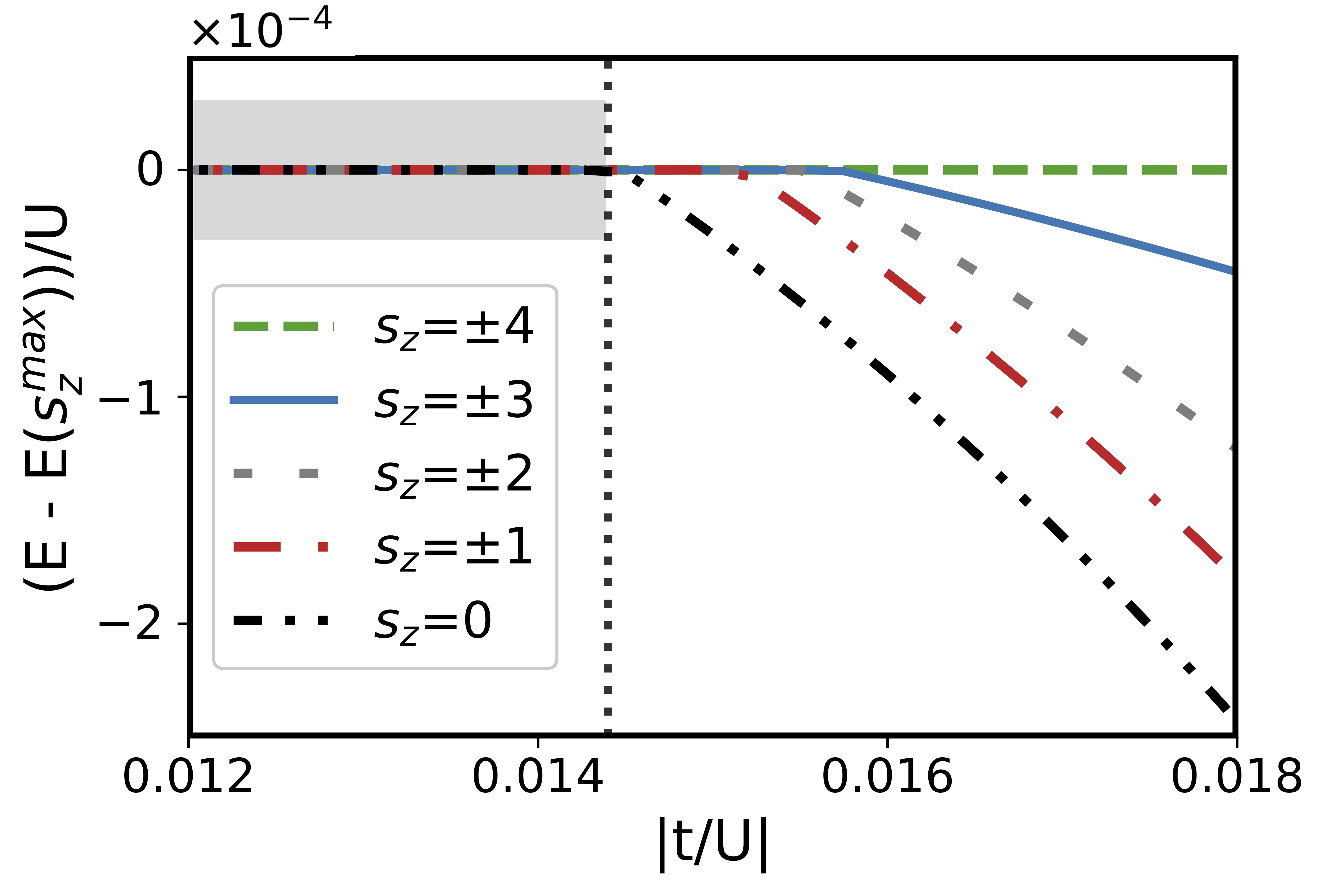}
    \caption{NF in a $3\times 3$ array. Black dotted line indicates the transition for NF. Ground state for each of the different spin $z$ subspaces plotted as a function of $|t/U|$, which are offset by the energy for $s_z = 4$. When the plotted energy difference for an $s_z$ ground state becomes negative, the state switches from maximum total spin $S=4$ to a lower total spin $S$.}
    \label{fig:3x3}
\end{figure}

We plot the ground state (GS) energy for each subspace in Fig. \ref{fig:3x3} (again shifted by the GS energy for maximum $s_z$). For a small $t/U$ or a large $U$, the $s_z$ states are all degenerate, i.e. the system has a maximum-total-spin ground state. This is in contrast to the usual non-ferromagnetic spin order near half-filling, where states have increasing energy as $s_z$ increases, as shown in Fig. \ref{fig:3x3}. The maximum spin GS holds up to $|t/U| \approx 0.0144$ (black dashed line). After this transition, it becomes higher in energy than the smaller total spin states and is no longer the ground state. Each $s_z$ GS other than the $s_z=4$ GS has a different total spin below and above the transition. 

\begin{figure}[h!]
    \includegraphics[scale=0.06]{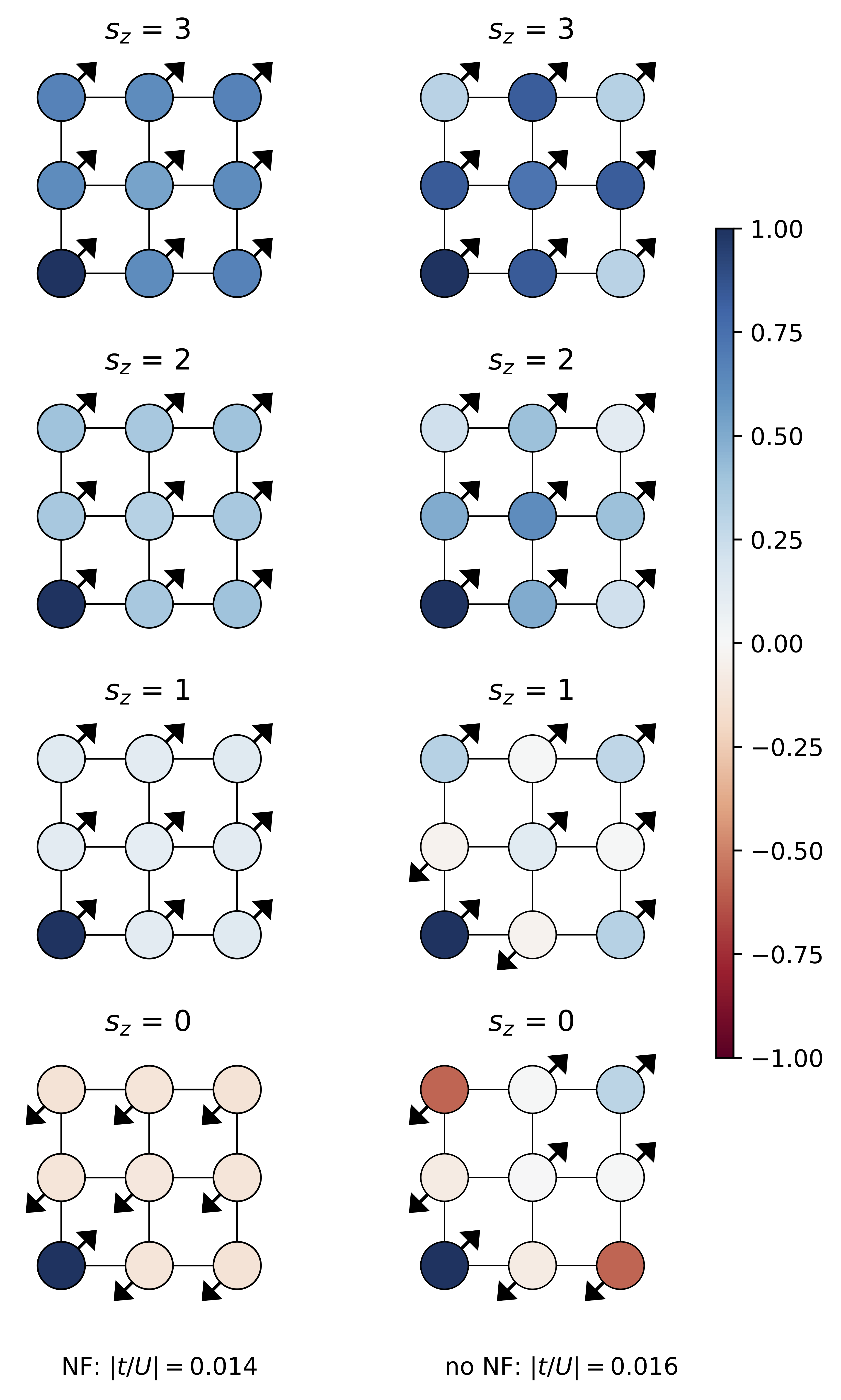}
    \caption{Ground state correlation function for the $3\times 3$ system below and above the NF transition when the lower-left array-site has a fixed spin-up electron. The correlation function for each non-negative spin-$z$ ($s_z= 3, 2, 1, 0$) ground state is shown. The negative $s_z$ GSs have the same plots but for correlation with a spin-down instead of a spin-up electron. The correlation function ranges from -1 to 1, indicated as the color ranges from dark red to dark blue. The arrows point in the upper right direction when the site is positively correlated, and to the bottom left when it is negatively correlated. In the NF regime, below the transition point, the ground state is ferromagnetic; above the transition, the ground state has a smaller total spin, the correlation plots show quasi-antiferromagnetic characteristics.}
    \label{fig:3x3_CF}
\end{figure}

In the $2\times 2$ array geometry the four sites are equivalent. The $3\times 3$ array geometry is more complicated than the $2\times 2$ geometry because there are three different, inequivalent kinds of sites: the four corner sites are like the $2\times 2$ sites with two connections only to other external sites; the four edge sites that have one additional hopping to the center site, and one center site that is the only internal site, with four nearest neighbor hoppings. Despite the more complicated structure, the energies for the $s_z$ GSs for the $3\times 3$ array follow a similar pattern to the GS energies of the $2\times 2$ array, as can seen by comparing Figs.~\ref{fig:2x2_sz} and \ref{fig:3x3}. The transition is at $|t/U|\approx 0.0144$, which is lower than the transition for the $2\times 2$ array at $0.053$, showing that larger interaction is needed for the transition to occur in the $3\times 3$ array. 

To see how electrons are organized below and above the transition, we calculate the pair correlation function (CF). The CFs are determined by fixing the number and spin of an electron on one chosen site and calculating the probability of having spin up/down electrons on the other sites. The probability of having spin up is subtracted from the probability for spin down, indicating whether the other site is positively/negatively correlated with the fixed site. The correlation function for a fixed configuration $\sigma_{f}$ on site $f$ with some other configuration $\sigma_{o}$ on site $o$ is, 
\begin{equation}
    C_{\sigma_{f},\sigma_{o}} = \frac{\sum \alpha^2[\sigma_{f}, \sigma_{o,\uparrow}]}{\sum \alpha^2[\sigma_{f}]} - \frac{\sum \alpha^2[\sigma_{f},\sigma_{o,\downarrow}]}{\sum \alpha^2[\sigma_{f}]}
\end{equation}
where $\sigma_{o, \uparrow / \downarrow}$ indicates the spin up/down occupation of site $o$, $\alpha[\sigma_{f}]$ is the probability amplitude for any configuration of site occupations that includes the $\sigma_{f}$ occupation, and $\alpha[\sigma_{f},\sigma_{o,\uparrow/\downarrow}]$ is the probability amplitude for any configuration that includes the site occupations at $f$ and $o$. The amplitudes' squared are summed over all configurations with the appropriate site occupations. The correlation function of the $s_z$ GS for the $3\times 3$ array when one spin-up electron is fixed on the corner is plotted in Fig. \ref{fig:3x3_CF}. The color indicates the sign and strength of the correlation, dark blue means the density on that site has more spin-up than spin-down and is positively correlated with the fixed up-spin at the corner site. A dark red site indicates the site has larger spin-down density. The direction of the arrow at each site also shows whether the site is positively or negatively correlated with the spin at the fixed site. Fig.\ref{fig:3x3_CF} compares the correlation function below and above the transition point for $s_z = 3, 2, 1, 0$. Correlations that are ferromagnetic below the transition becomes locally more antiferromagnetic above the transition for all $s_z$ spaces.

\begin{figure}[h]
    \includegraphics[scale=0.07]{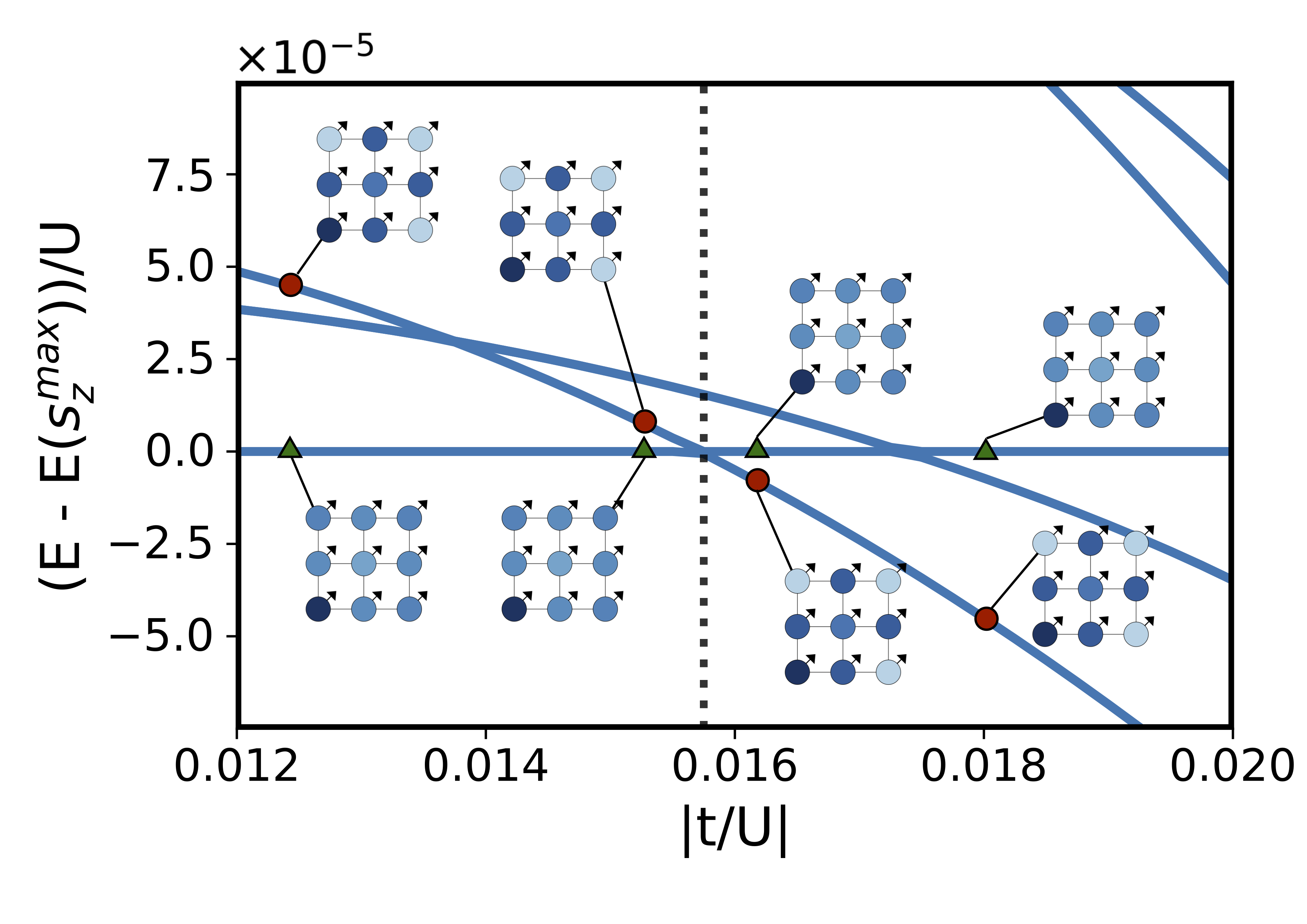}
    \caption{Correlation function and energies for $s_z = 3$ as a function of $|t/U|$ near the transition point. The energies are offset by the $s_z=4$ state energy. A spin-up electron is fixed at the lower-left array-site. Correlation plots are displayed at marked energies. Green triangles indicate a ferromagnetic state ($s_z=3$, $S=4$), while red circles indicate a state with a smaller total spin ($s_z = 3$, $S=3$). The black dotted line is the transition point. Note that for $s_z = 3$ subspace, the transition point is larger at $|t/U| \approx 0.0157$ instead of $0.0144$. Markers are $t/U$ values chosen arbitrarily to show how correlation plots change ($|t/U| = 0.0125, 0.0153, 0.0162, 0.018$).}
    \label{fig:3x3_follow}
\end{figure}

From the correlation plots shown in Figs. \ref{fig:3x3_CF}, the states below and above the transition look qualitatively different because the GS for each spin $z$ switches to a state with the same spin $z$ but smaller total spin $S$. This is confirmed by following the first few excited states through transition. Fig. \ref{fig:3x3_follow} shows a blow-up of the region near the transition. Below the transition, the ground state is the maximum-total-spin state. An excited state below the transition with the same $s_z$ but smaller total $S$ becomes the ground state above the transition. As $|t/U|$ moves away from the transition point, the correlation functions for different total-spin states do not change qualitatively, but vary slowly and continuously across and away from the transition.

Nagaoka ferromagnetism occurs when there is one hole in a half-filled band. We have checked for other fillings. The one-hole case is the only filling that exhibits saturated ferromagnetism with a maximum-total-spin ground state at large enough $U$.

\subsection{\label{NxN}$N\times N$ arrays}
NF was first predicted for a 2D extended bulk system of electrons with infinite interaction $U$. The bulk limit should be reached as the size of an $N\times N$ array increases. To see if NF for one hole in a half-filled band still occurs for small $t/U$ as $N$ increases and, if so, how the transition point changes as the array size increases, and to see if ferromagnetism can appear for smaller band fillings for larger $N$ we have studied $N\times N$ arrays for $N$ = 4, 5, 6, 7 and 8.  

\begin{figure}[h]
\subfloat[$4\times 4$ and $5\times 5$]{%
  \includegraphics[scale=0.026]{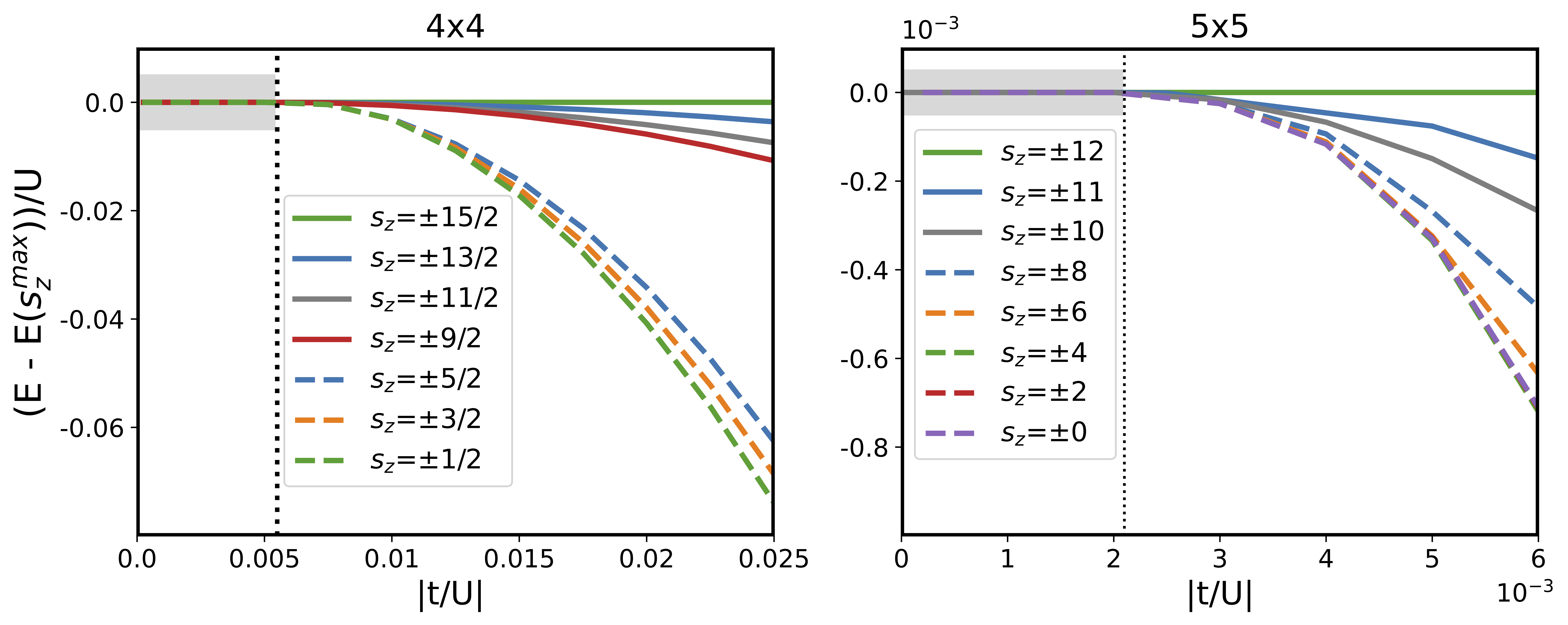}%
  \label{fig:NxN}
}\\
\subfloat[$N\times N$]{%
  \includegraphics[scale=0.052, left]{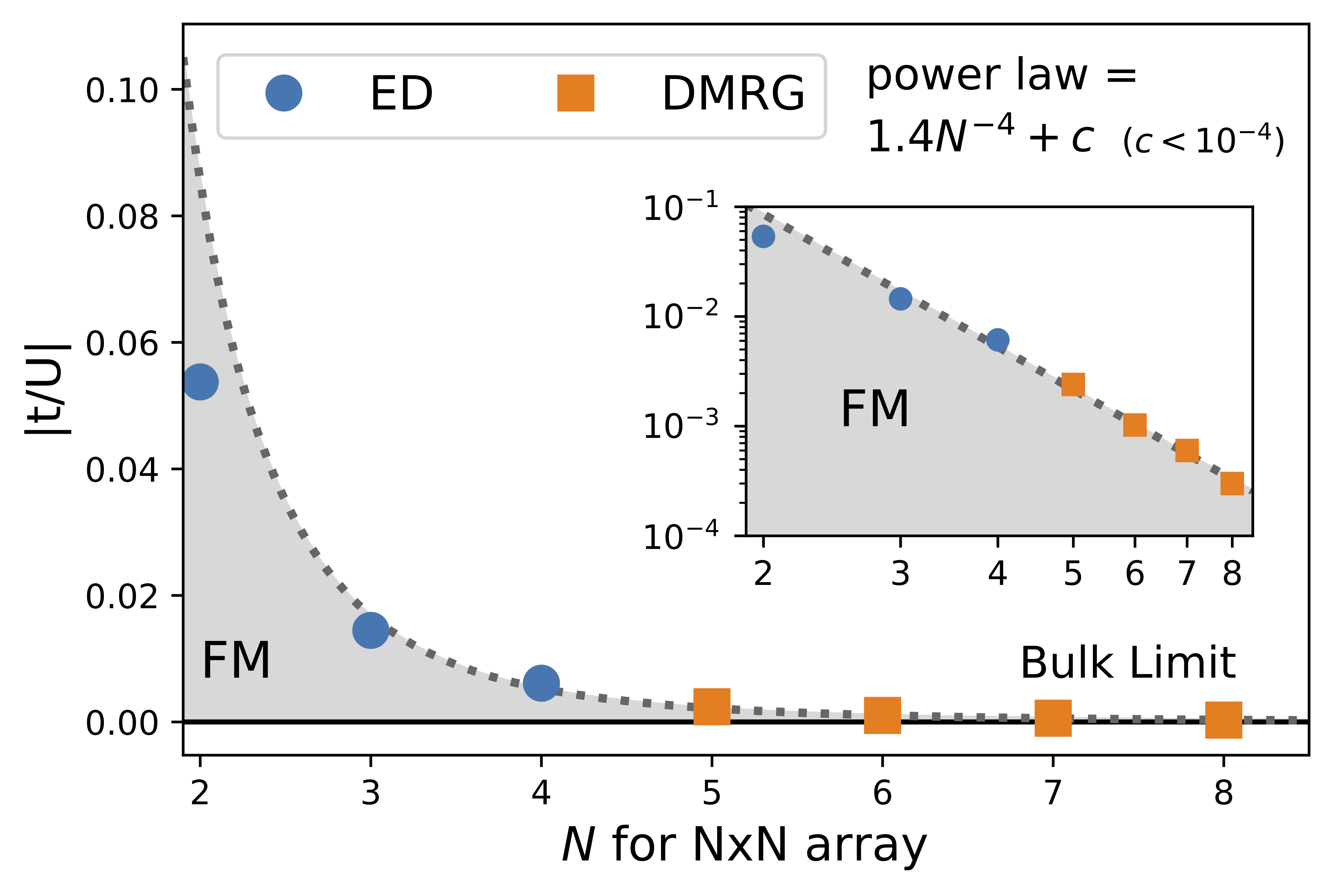}%
  \label{fig:NxN_en}
}
\caption{NF for $N\times N$ arrays. (a) NF for $4\times 4$ and $5\times 5$ arrays. Energy of GS for each $s_z$ subspace plotted as a function of $|t/U|$, offset by the energy of the maximum $s_z$ state. The black dashed line indicates the transition for NF. Both system are in the one hole case, $4\times 4$ arrays have $S_{max}^{4\mathrm{x}4}=15/2$, and for $5\times 5$ arrays $S_{max}^{5\mathrm{x}5} = 12$. Here $s_z$ subspaces accessible with ED are shown as solid lines, DMRG as dashed lines. (b) Transition point v.s. $N$, including both ED and DMRG results. Fitted power law functions (discarding lower values of N) shows how the system approaches the bulk limit (Nagaoka limit $U \rightarrow \infty$) as $N$ gets larger. Inset axis shows the plot in log-log scale.}
\label{fig:sq_arr}
\end{figure}

Simulation done for $4\times 4$ and $5\times 5$ arrays for one-hole in a half-filled band are shown in Fig.\ref{fig:NxN}. The larger $N\times N$ arrays behave similarly to the $3\times 3$ and $2\times 2$ arrays, with the transition point decreasing as N gets larger, as shown in Fig. \ref{fig:NxN_en}. As the $N\times N$ array gets larger, the on-site interaction $U$ needs to be larger for NF to occur. This agrees with Nagaoka's original theorem, that ferromagnetism in an infinite lattice occurs for infinite on-site interaction $U$. Fig. \ref{fig:NxN} includes the results from ED and density matrix renormalization group (DMRG) simulations. For smaller $s_z$ the large number of possible spin configurations makes ED difficult even for $4\times 4$ and $5\times 5$ arrays. When ED calculations are not possible, we perform DMRG simulations. Here, we flatten the 2D arrays into 1D chains with long-range hoppings, and obtain their ground state energies through the well-established two-site iterative variational searches \cite{ITensor, ITensor-r0.3}. The DMRG calculations confirm the prediction that for larger $N\times N$ arrays, NF exists for a smaller region of $|t/U|$. The dependence of the transition ratio on array size $N$ shown in Fig. \ref{fig:NxN_en} out to the $8\times 8$ array has been well fitted to an inverse power law on $N$. In the Appendix \ref{scaling}, a scaling argument is used to support an inverse power law dependence on $N$. The $7\times 7$ is the first lattice with more internal sites than edge sites. Our results suggest that the bulk limit for NF ferromagnetism with $U$ becoming very large is being approached by $N=8$. Table~\ref{tab:NxN} summarizes the results for $N\times N$ arrays. Previous work on the $9\times 9$ array with open boundary conditions in \cite{white2001} suggests that the transition point to a ferromagnetic ground state on the whole system is $J/t = 4t/U \approx 0.001$, which is slightly larger than the transition point we predict here from scaling from the $8\times 8$ array to the $9\times 9$ array.

\begin{table}[h!]
\begin{ruledtabular}
\begin{tabular}{cccccc}
Geometry & Sketch & NF & $\#$ of $e^-$& Transition & Methods\\ \hline
$2\times 2$ & \includegraphics[height=0.3in]{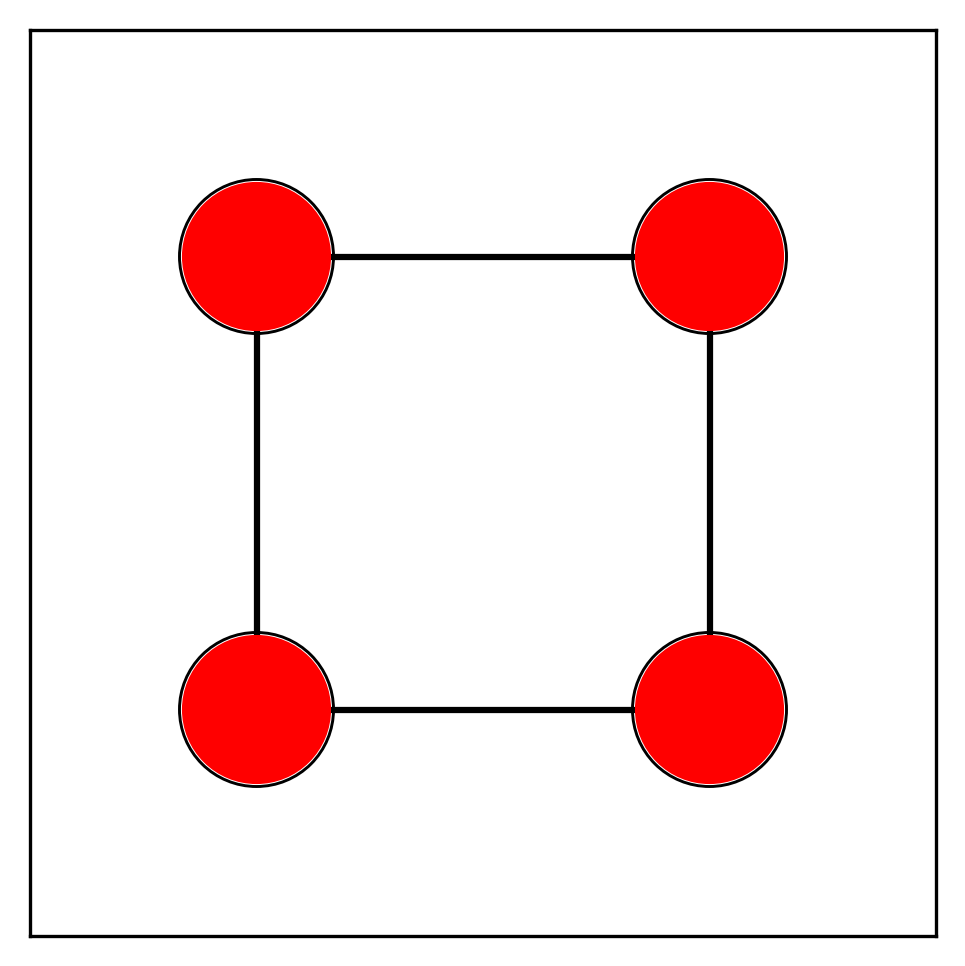} & yes & 3 & $0.053$ & ED \\ \hline
$3\times 3$ & \includegraphics[height=0.3in]{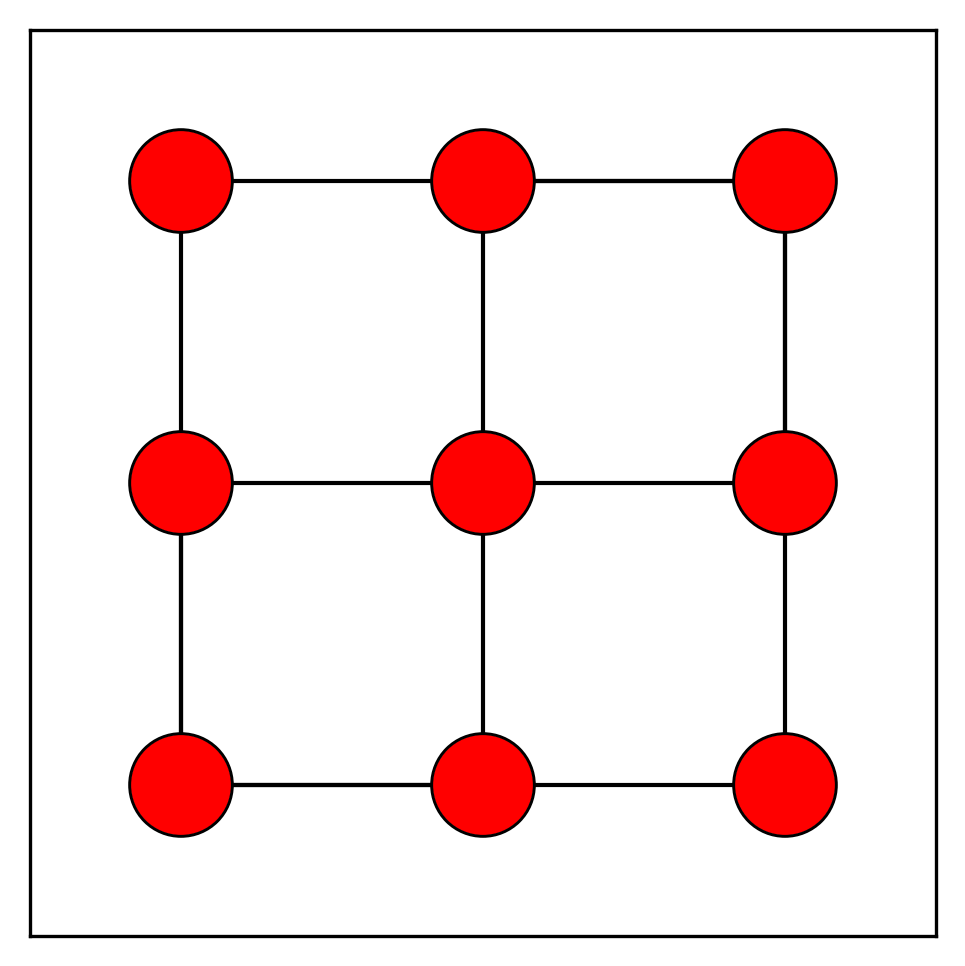} & yes & 8 & $0.0144$ & ED \\ \hline
$4\times 4$ & \includegraphics[height=0.3in]{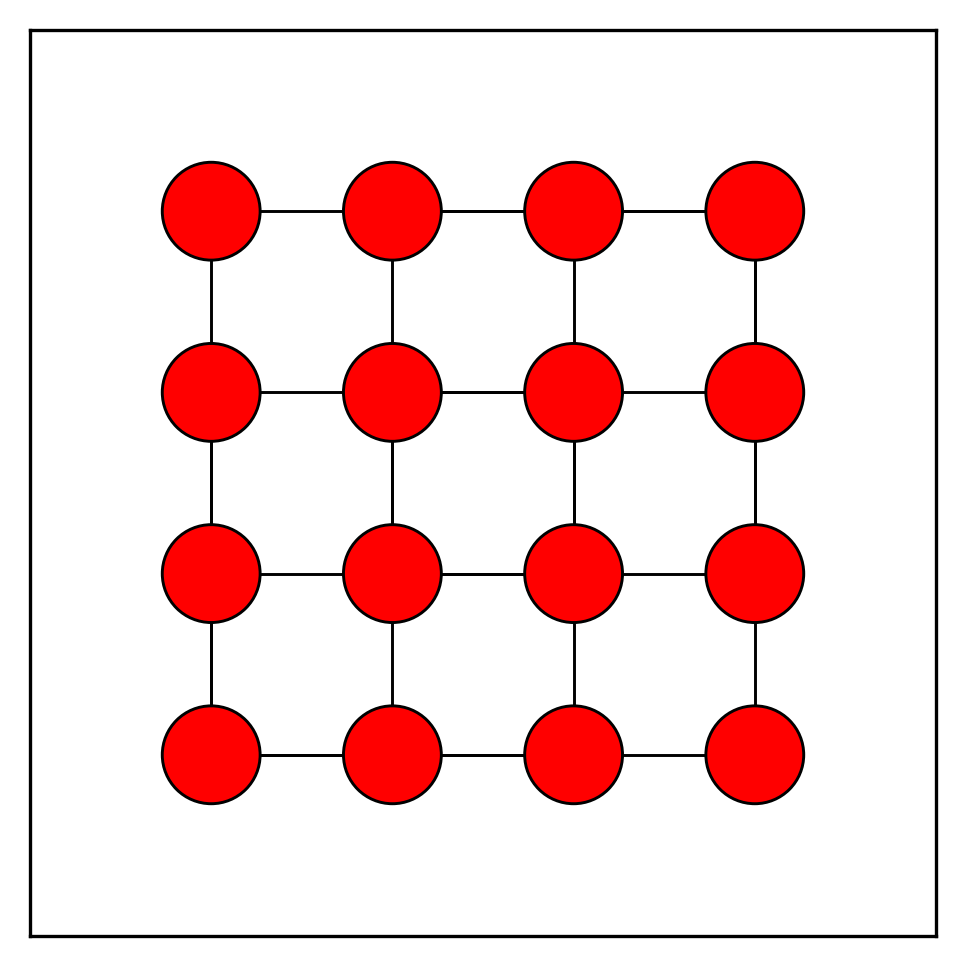} & yes & 15 & $0.0061$ & ED, DMRG \\ \hline
$5\times 5$ & \includegraphics[height=0.3in]{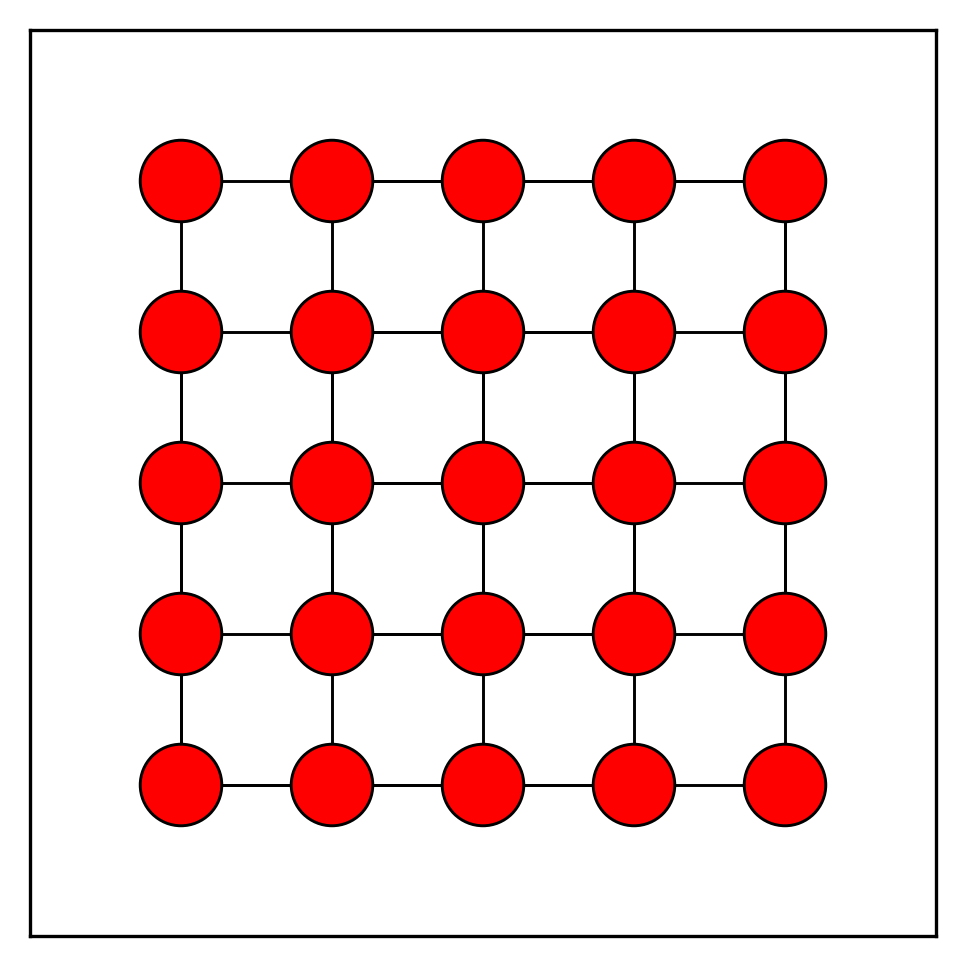} & yes & 24 & $0.0024$ & ED, DMRG \\ \hline
$6\times 6$ & \includegraphics[height=0.3in]{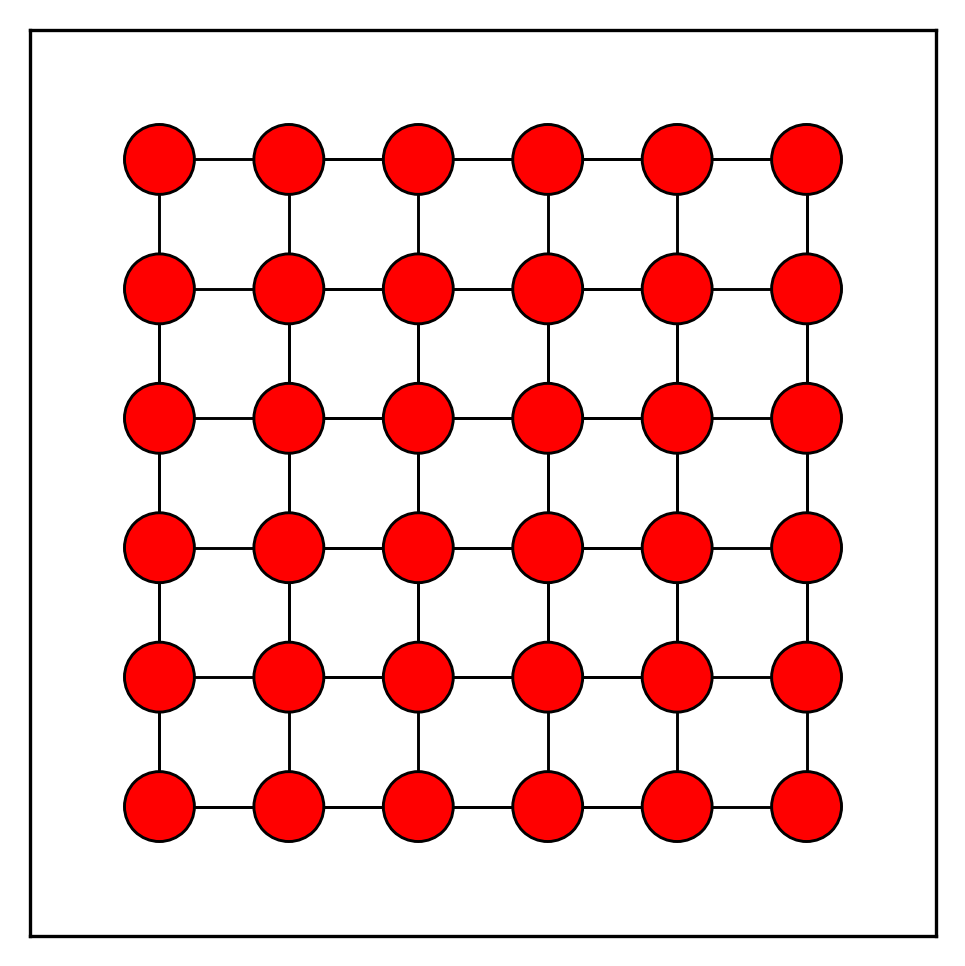} & yes & 35 & $0.001$ & DMRG \\ \hline
$7\times 7$ & \includegraphics[height=0.3in]{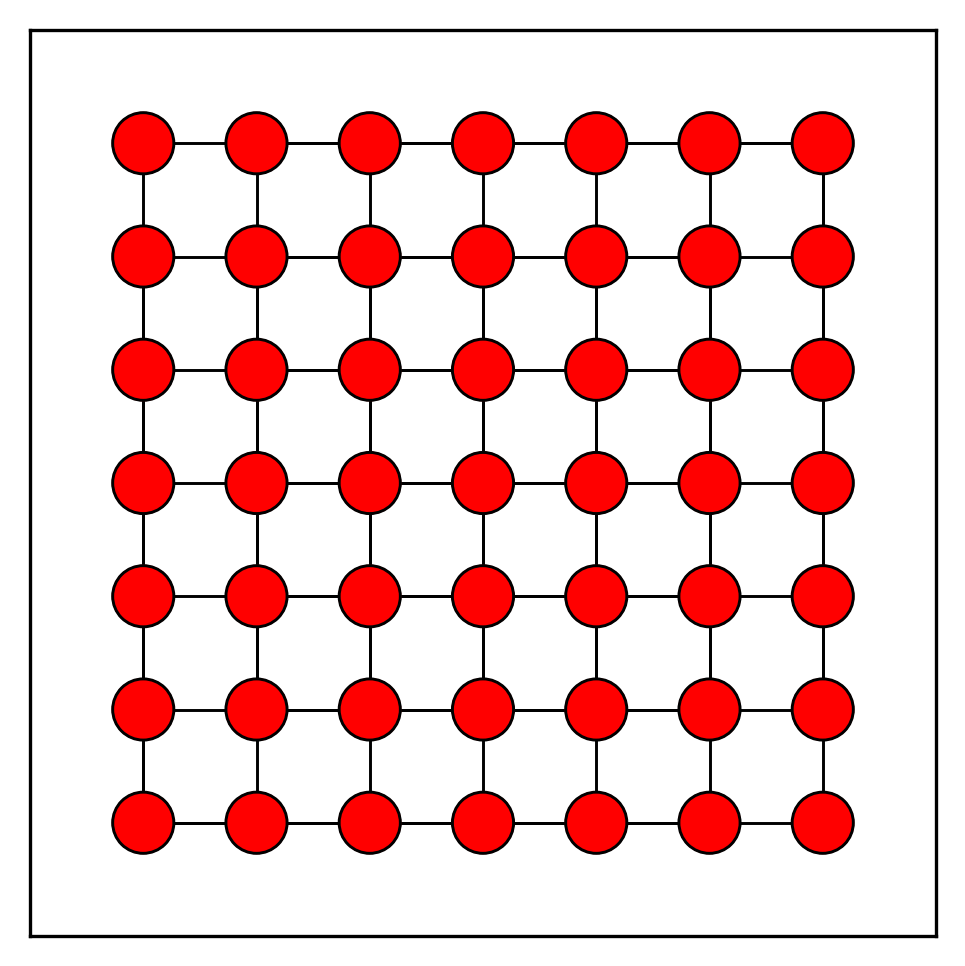} & yes & 48 & $0.0006$ & DMRG \\ \hline
$8\times 8$ & \includegraphics[height=0.3in]{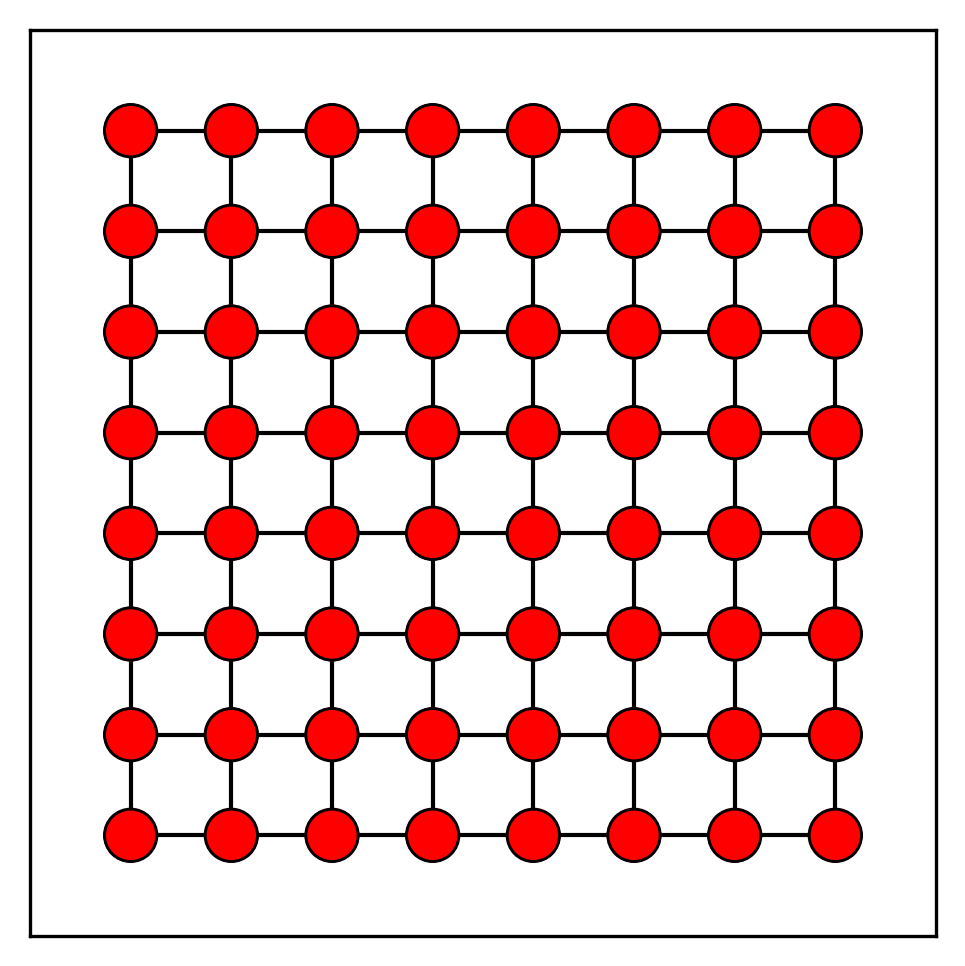} & yes &63 & $0.0002$ & DMRG \\
\end{tabular}
\end{ruledtabular}
\caption{\label{tab:NxN}$N\times N$ arrays results: geometry, sketch, whether Nagaoka ferromagnetism exists, number of electrons, transition point, and calculational methods.}
\end{table}
 
\section{\label{robustness}Different Geometries beyond the $3\times 3$ Arrays}  

 The sites in a semiconductor array need not be identical. The site energies can be controlled by bias. In dopant systems, disorder in dopant position and dopant number can affect the tunnelling and on-site energies. For example, there can be more than one dopant clustered near a site. The properties of that disordered site will be determined by the multidopant cluster which will have a different on-site energy. It is harder for electrons to hop on and off a disordered site with a shifted on-site energy, lowering the participation of that site in the array states. When the energy of the disordered site is sufficiently far off-resonance from the other sites, the disordered site is effectively removed from the array. To investigate the robustness of NF in fabricated $3\times 3$ arrays with disordered sites, we consider $3\times 3$ arrays with one site missing to see if ferromagnetism survives or disappears, and, if ferromagnetism survives, what electron fillings display the ferromagnetism. This will provide us insight into how and when NF exists in small finite systems with variations in on-site energies. In this paper we look at the limit when the on-site energy difference is large enough to remove the site. In an upcoming paper, we will show how the ferromagnetism evolves as on-site energy differences vary. 

\begin{figure}[h!]
\subfloat[]{%
  \includegraphics[scale=0.25]{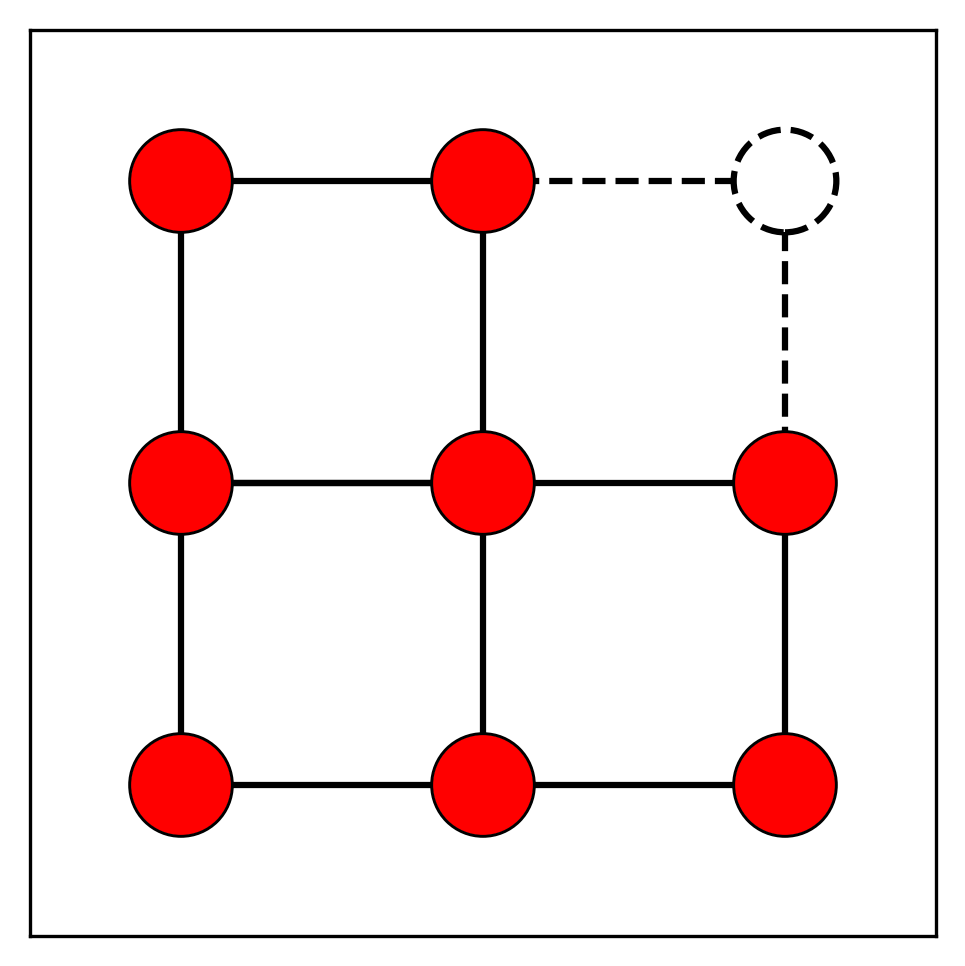}%
  \label{fig:nocor}
}\hfill
\subfloat[]{%
  \includegraphics[scale=0.25]{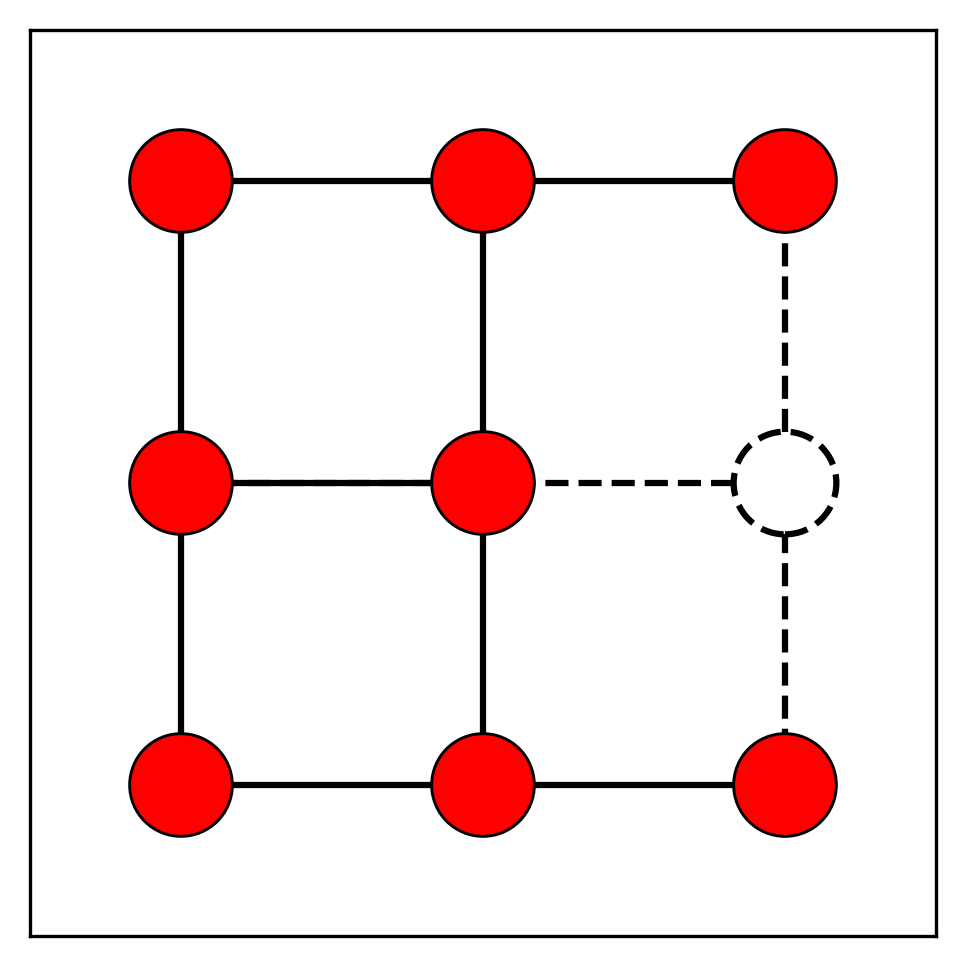}%
  \label{fig:noedg}
}\hfill
\subfloat[]{%
  \includegraphics[scale=0.25]{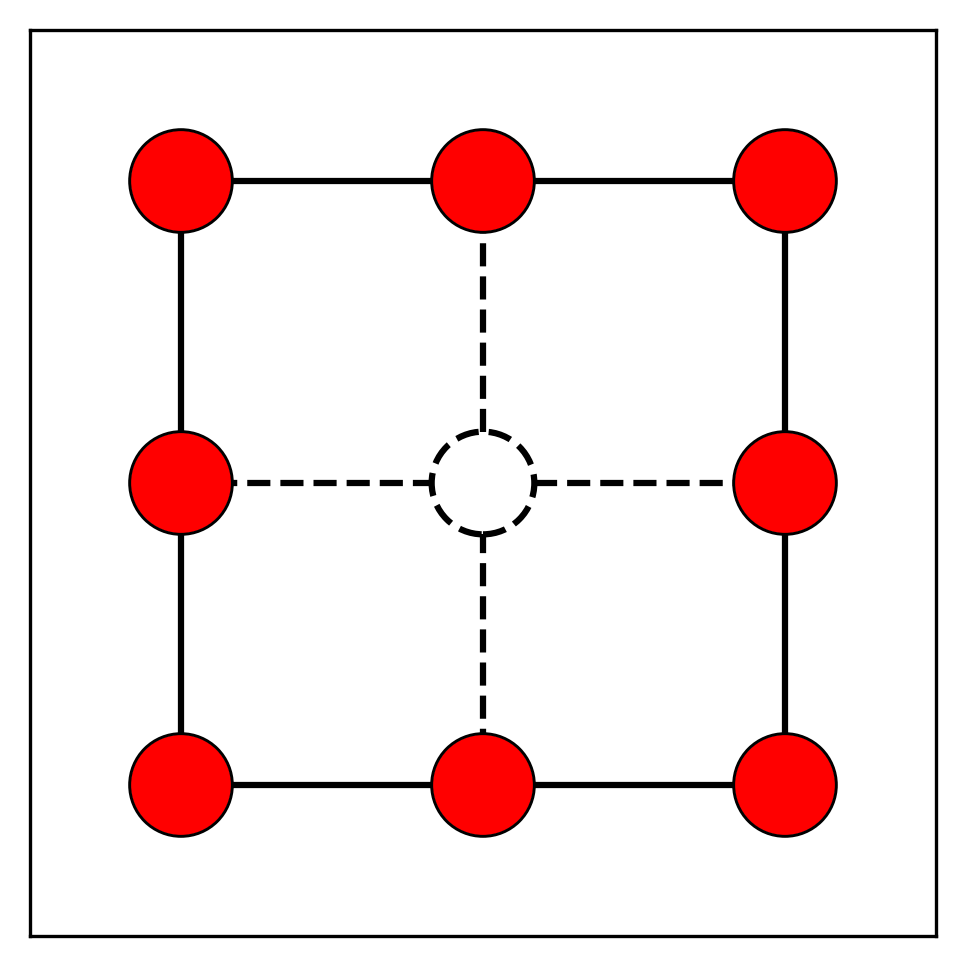}%
  \label{fig:nocen}
}
\caption{Three ways of removing one site from a $3\times 3$ array at (a) a corner, (b) an edge and (c) the center. }
\label{fig:missing_atom}
\end{figure}

For the $3\times 3$ array, there are three different types of site, the corner, edge, and center sites, corresponding to having 2, 3, or 4 hoppings connecting it to other other sites. Removing one site of each type of atom is shown in Fig. \ref{fig:missing_atom}. We discuss these three cases in the next three sections. We also discuss ferromagnetism in other finite structures that can be made by removing similar sets of sites in larger $N\times N$ arrays. We summarize the results for removing sites from the $3\times 3$ array and other extended structures in Table \ref{tab:robustness}.

\begin{table}
    \centering
    \begin{ruledtabular}
    \begin{tabular}{p{1.5cm}p{1.3cm}p{0.9cm}p{0.7cm}p{1.5cm}p{1.5cm}}
        Geometry & Sketch & NF/IF? & $\# e$& Transition & Details\\ \hline
        \makecell{$3\times 3$} & \includegraphics[height=0.4in]{3by3.png} & NF  & 8 & $0.053$ & 1-hole NF \\ \hline
        \makecell{no corner} & \includegraphics[height=0.4in]{nocor.png} & NF  & 7 & $0.0104$ & 1-hole NF  \\ \hline
        \makecell{no edge} & \includegraphics[height=0.4in]{noedg.png} & no &  &  & not connected \\ \hline
        \makecell{no center} & \includegraphics[height=0.4in]{3loop.png} & IF & 3 & $0.131$ & $3e$ IF \\ \hline
        \makecell{no same-side corner} & \includegraphics[height=0.4in]{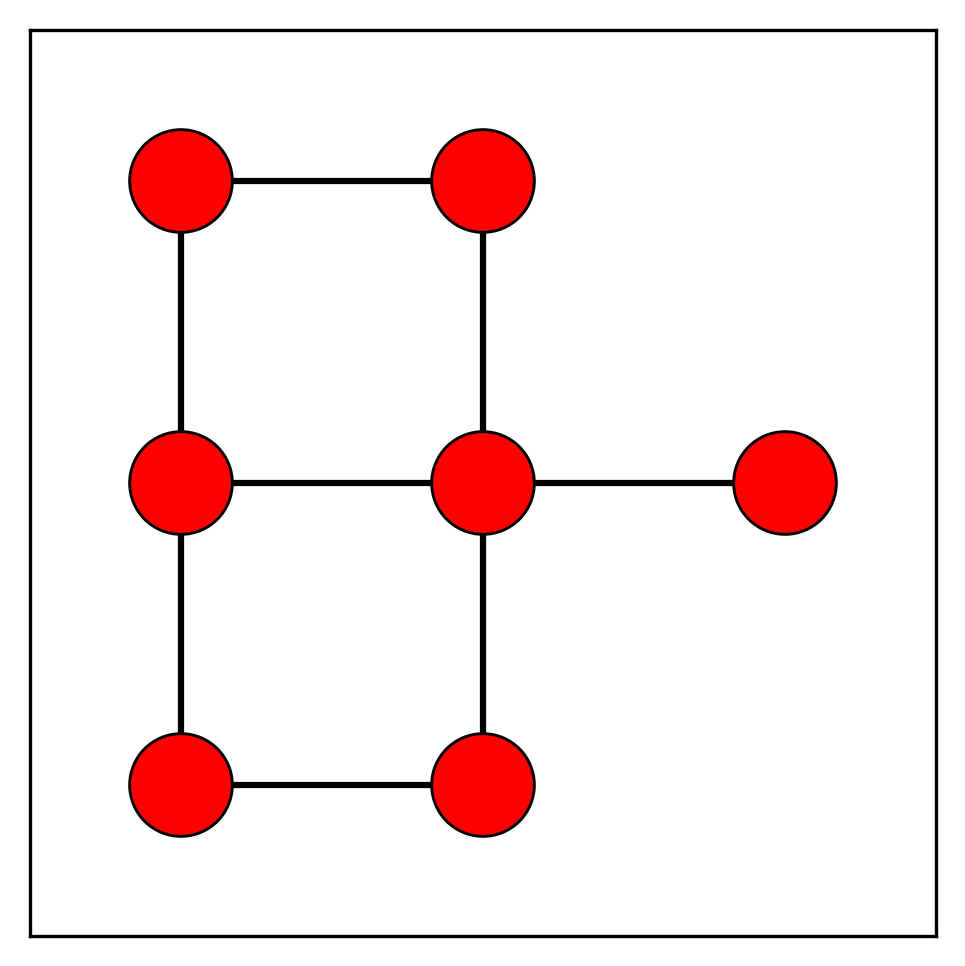} & no  &  &  &  not connected \\ \hline
        \makecell{no opposite corner \\ (2 corner-sharing sq)} & \includegraphics[height=0.4in]{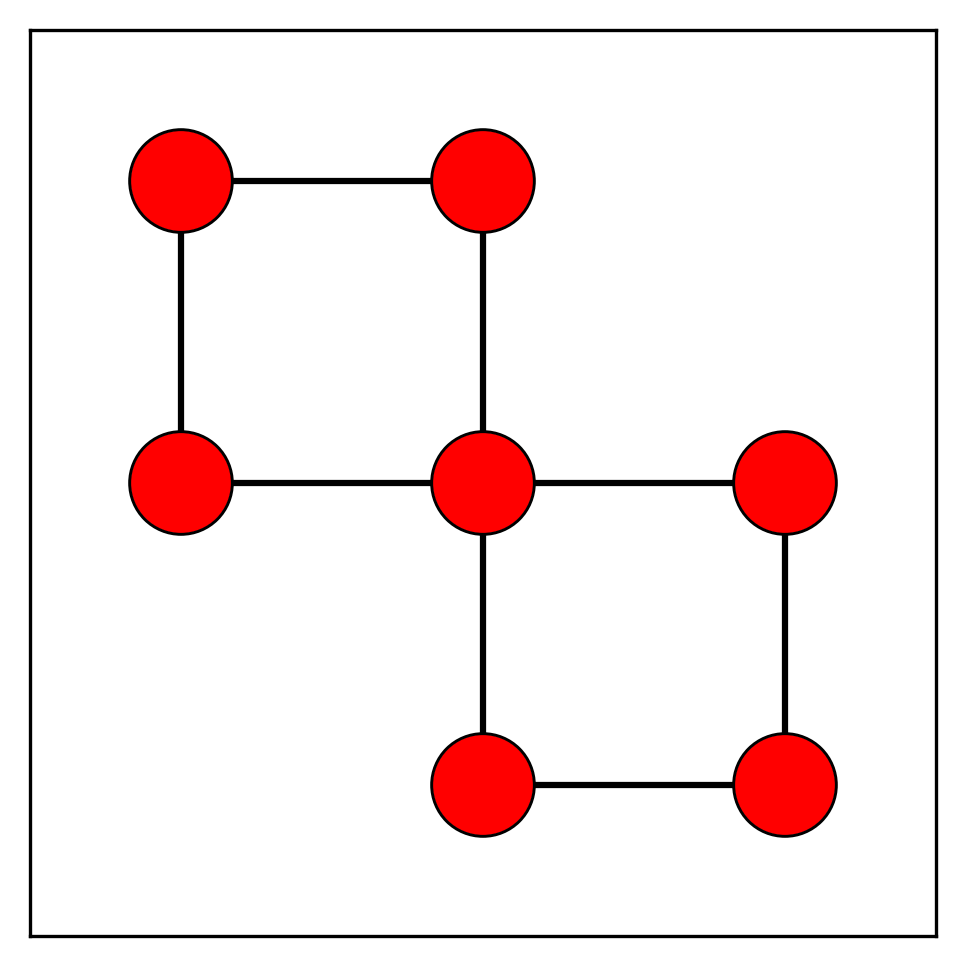} & IF  & 5 & $0.0127$ & 2-hole IF \\ \hline
        \makecell{2x3 \\ (2 side-sharing sq)} & \includegraphics[height=0.4in]{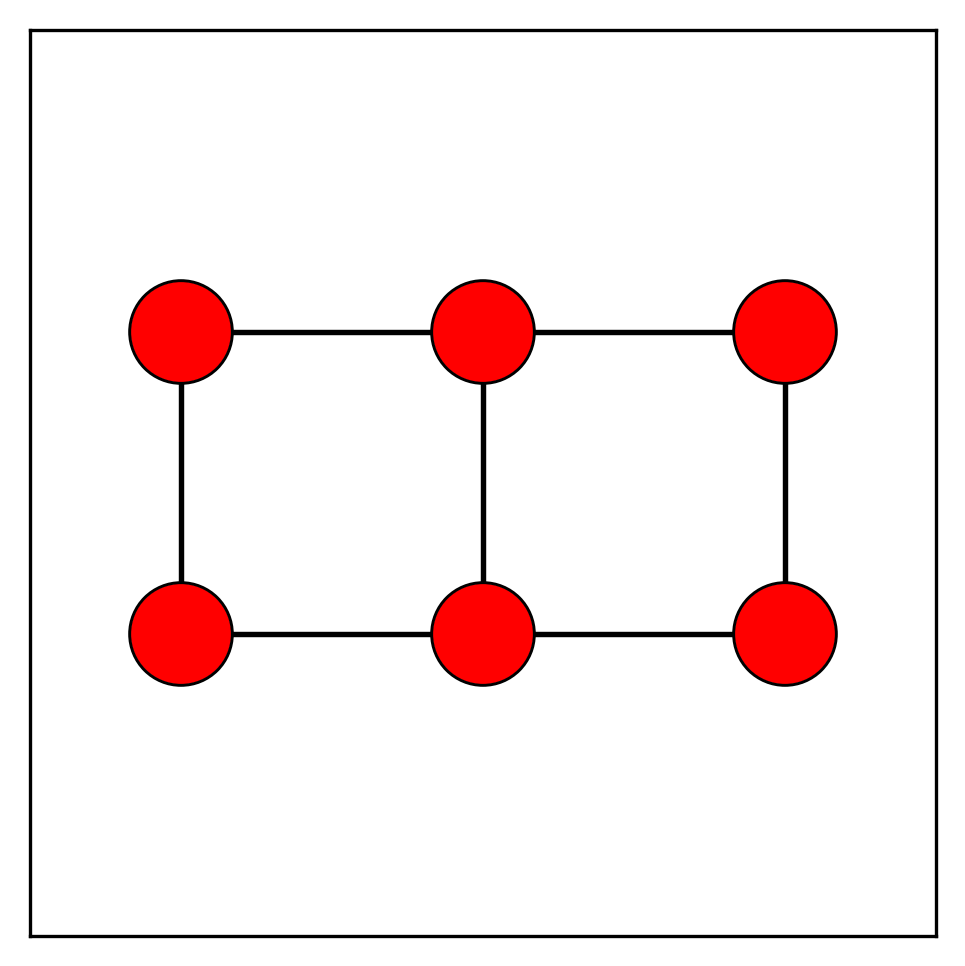} & NF  & 5 & $ 0.019$ & 1-hole NF \\ \hline
        \makecell{$N$-site loops} & \includegraphics[height=0.4in]{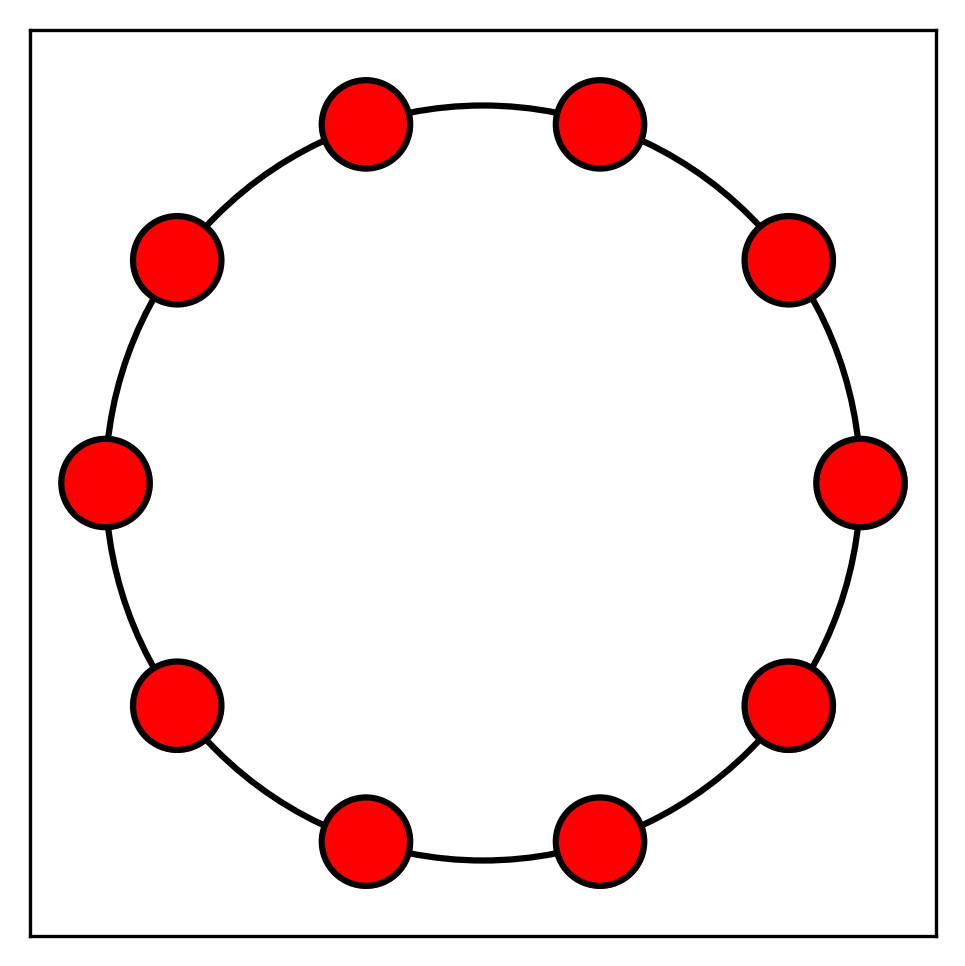} & IF  & 3 & \makecell{$0.017N$ \\$-0.002 $} & $3e$ IF \\
        
    \end{tabular}
    \caption{Summary table for other geometries, their sketch, whether Nagaoka ferromagnetism (NF) or itinerant ferromagnetism (IF) exists, number of electrons, and transition point if applicable.}
    \label{tab:robustness}
    \end{ruledtabular}
\end{table}

\subsection{\label{nocor}Removing Corner Atoms - Rectangular Arrays} 
\subsubsection{Removing One Corner Atom}
We consider first removing one corner atom (shown in Fig. \ref{fig:nocor}), which removes the fewest hoppings and gives a structure most similar to the original $3\times 3$ array. After one of the corner atoms is removed, the structure becomes three $2\times 2$ squares stacked together, sharing common sides. NF still exists but now for one-hole in the new half-filled band, i.e. now for 7 electrons. The transition point at $|t/U| \approx 0.0104$ is slightly lower than for the full $3\times 3$ array at $0.0144$. Removing one site can be modeled by increasing the on-site energy of that site. When the on-site energy is zero (same as the other sites in the array), the system is the full $3\times 3$ array. When the on-site energy is infinite, the array has one less site. In this paper, we discuss these two limits for the on-site energy of the disordered site. In an upcoming paper, we will show how the ferromagnetism evolves from NF on a $3\times 3$ array to NF on an array with one less corner site and one less electron as the energy shift of the disordered corner site is increased. 

\subsubsection{Removing Two Corners on the Same Side}
There are two different ways to remove two corner sites in a $3\times 3$ array, as shown in lines 5 and 6 in Table \ref{tab:robustness}. If two corner atoms on the same side are removed, the structure becomes a $2\times 3$ rectangle connected by one extra hopping to an edge site. This geometry does not exhibit GS ferromagnetism for one hole in the half-filled band. 

\subsubsection{Role of Wavefunction Connectivity}
\begin{figure}[h!]
\subfloat[]{%
  \includegraphics[scale=0.27]{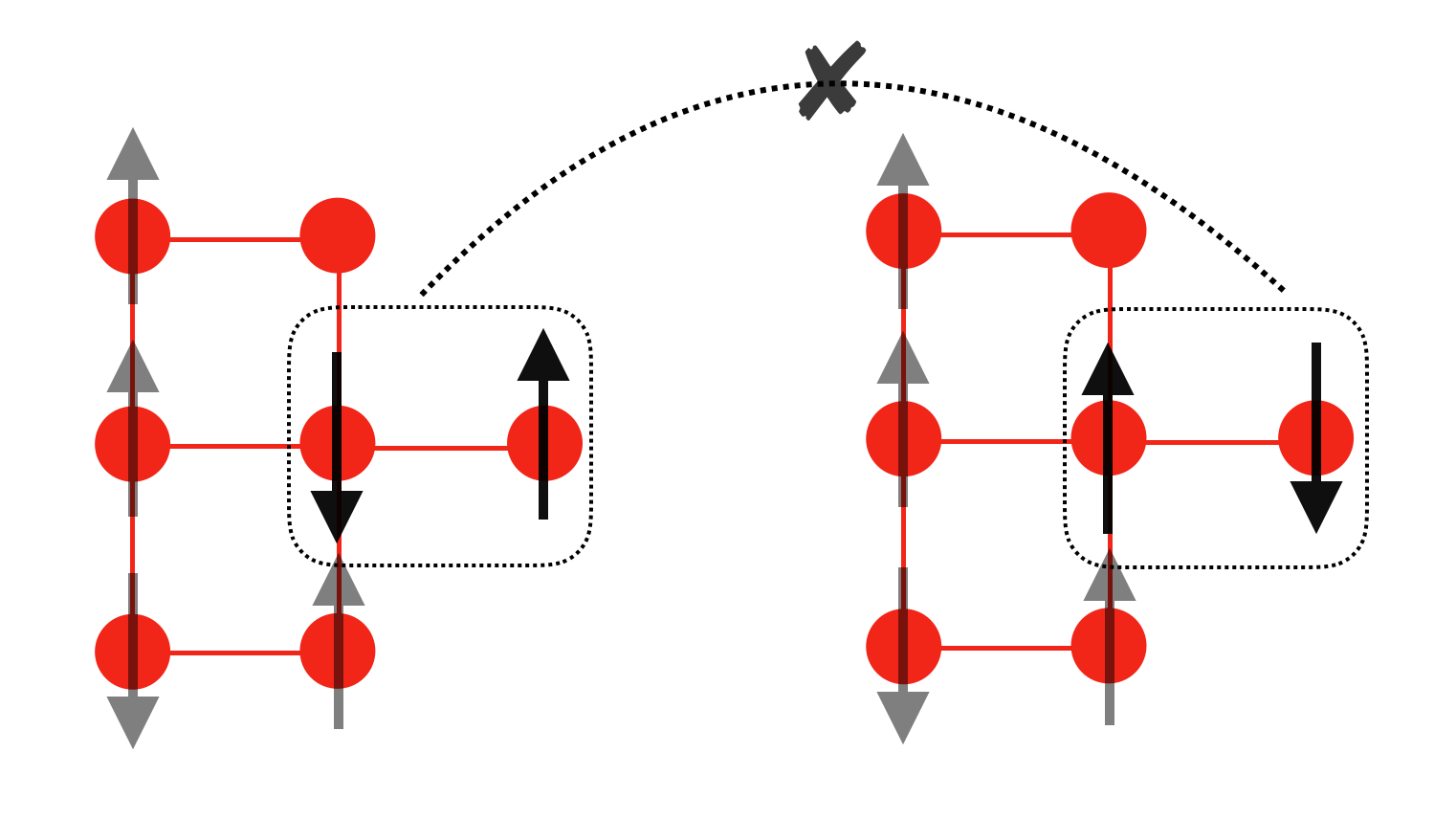}%
  \label{fig:not_con}
}\\
\subfloat[]{%
  \includegraphics[scale=0.4]{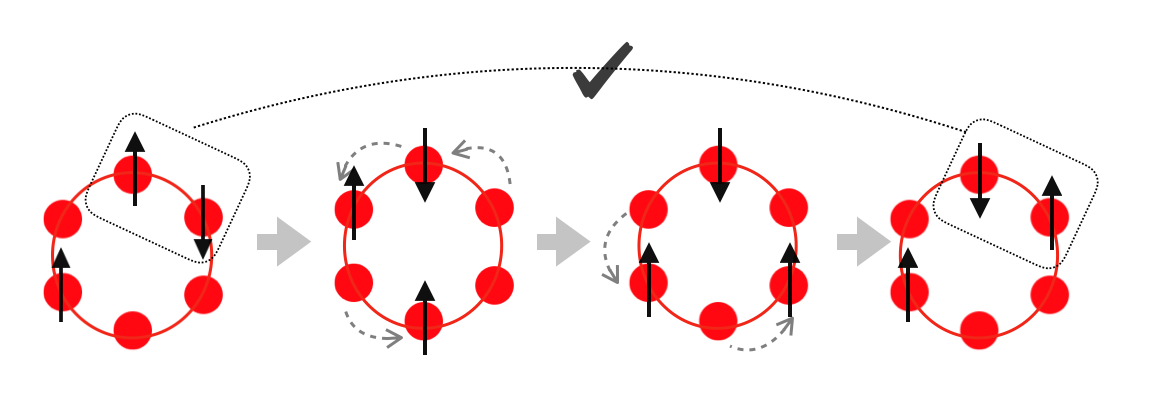}%
  \label{fig:connected}
}
\caption{(a) An example showing when the connectivity condition for NF is violated. To exchange the two spins (in the box), a series of nearest neighbor hoppings is not sufficient, as the spin up and spin down electrons must be on the same site which is strongly suppressed by large repulsion $U$. (b) Example of the connectivity condition being satisfied for 3 electrons on a loop. For any sized loop, 3 electrons is the largest filling where the wavefunction is connected. Any pair of spins (shown in the box) can be exchanged when all electrons are hopped around the loop.}
\label{fig:connectness}
\end{figure}

Tasaki showed \cite{tasaki1998} that for NF to exist the many-body wavefunction must satisfy a connectivity condition. Here, being a connected many-body wavefunction requires that each site be connected to other sites in the lattice by tunnel coupling and, more importantly, that any configuration of spins in the array can be connected, via hopping without double occupancy, to any other configuration of the same spins in the array. Connectivity is a necessary but not sufficient condition for saturated ferromagnetism. For an array with a hanging single hopping as shown in line 5 of Table \ref{tab:robustness}, a spin on the isolated site connected to the rest of the array by one hopping can be moved off the isolated site only by exchanging with a hole on its only neighbor. However, the spin can't be moved further to another site in the array unless a second hole is present. This is shown in Fig. \ref{fig:not_con}. The many-body wavefunction for this geometry is not connected for one-hole in a half-filled band and can't realize saturated ferromagnetism for that filling. This prohibits NF for one-hole in a half-filled band. However, the many-body wavefunction becomes connected when there are two or more holes in the half-filled band. However, we don't find GS saturated ferromagnetism for any number of holes. This shows that the connectivity condition is necessary but not sufficient.

\subsubsection{Removing Two Opposite Corners }
If the two sites are removed from opposite corners of the $3\times 3$ array, the structure becomes two $2\times 2$ plaquettes connected by sharing a common corner site. The ground state for this structure is ferromagnetic only when there are two holes in a half-filled band of the array, that is, when there are five electrons. The transition is at $|t/U| \approx 0.0127$ and calculations of the pair correlation function show ferromagnetic behavior below the transition. We refer to this as the itinerant ferromagnetism (IF) case because it is not possible for one-hole doping of a half-filled band as in NF. This ferromagnetism for two-hole doping in a half-filled band can be extended to longer chains of squares. The ground state of the longer chains of corner-sharing squares is ferromagnetic but only when there are two holes in a half-filled band, and not ferromagnetic for the one-hole case (the NF state). The many-body wavefunction is not connected when there is one-hole in a half-filled band, preventing GS saturated ferromagnetism. The site shared by the two $2\times 2$ plaquettes acts as a bottleneck. A hole on one of the $2\times 2$ plaquettes can't be used to move a spin around another plaquette that has spins at each site. A second hole is needed. Hence ferromagnetism is possible for two holes, as we have found. As before, this is a necessary but not sufficient condition. For different finite chains of squares calculated here, saturated ferromagnetism does not exist for more than two holes. Much longer chains ($N > 100$) with periodic/open boundary conditions and finite hole doping densities are discussed in \cite{monten2006, monten2014}.
As the chain gets longer, the transition point also decreases as happens when arrays increase in size as seen in Fig \ref{fig:2-hole}. (For details, see Appendix \ref{2-hole}).
\begin{figure}[h!]
    \centering
    \includegraphics[scale=0.05]{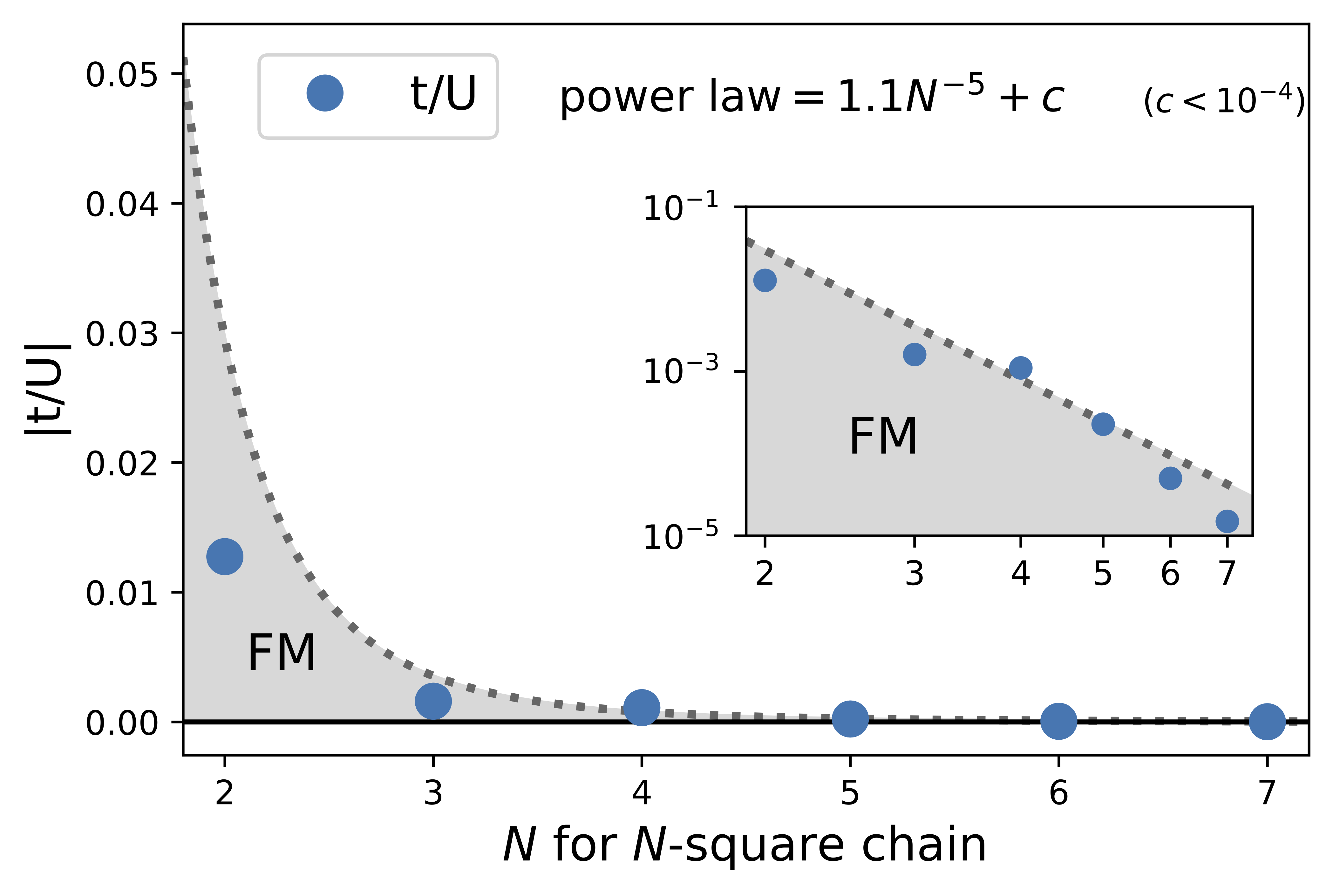}
    \caption{Two-hole ferromagnetism in a chain with $N$ $2\times 2$ plaquettes sharing corners on a common diagonal. Transition point versus the number $N$ of plaquettes. The fitted power law function is shown as a dotted line, fitting to only larger values of N = 5, 6, 7. Inset axis shows the log-log scale.
}
    \label{fig:2-hole}
\end{figure}

One might think that the 2-hole ferromagnetism occurs when the two holes are localized to the two $2\times 2$ plaquettes at the end of the chain. However, calculations of the local charge density show that the two holes are spread across the entire chain. The two-hole condition appears to arise primarily from the wavefunction connectivity.

\subsection{\label{noedg}Removing an Edge Site - One-fold Coordinated Sites}
When the middle site on an edge is removed, here referred to as an edge site, the array has two corner sites attached to a $2\times 3$ rectangle by single bonds. The corner sites are now one-fold coordinated. For this array, there is no ferromagnetism for any number of holes in the half-filled band, i.e. for any electron filling. As seen for an array with same-side corners removed, the many-body wavefunction is not connected for one-hole doping of the half-filled band of the array with a missing edge site. Thus, NF is not possible. The many-body wavefunction becomes connected for two holes, but there is still no GS ferromagnetism for any number of holes.

If the single hoppings and the isolated sites are removed, the structure turns into a rectangle with internal side-to-side hoppings, and NF appears again for one hole in the half-filled band of the rectangle. The rectangle can be extended to longer chains of $2\times 2$ plaquettes with each pair of adjacent squares sharing a common side. Again, the transition point decreases as the system gets longer. The results for the rectangles are discussed in Appendix \ref{other-NF}.

\subsection{\label{nocen}Removing the Center Site - Loops}
\subsubsection{Three-electron Loop Ferromagnetism}
\begin{figure}[h]
    \includegraphics[scale=0.045]{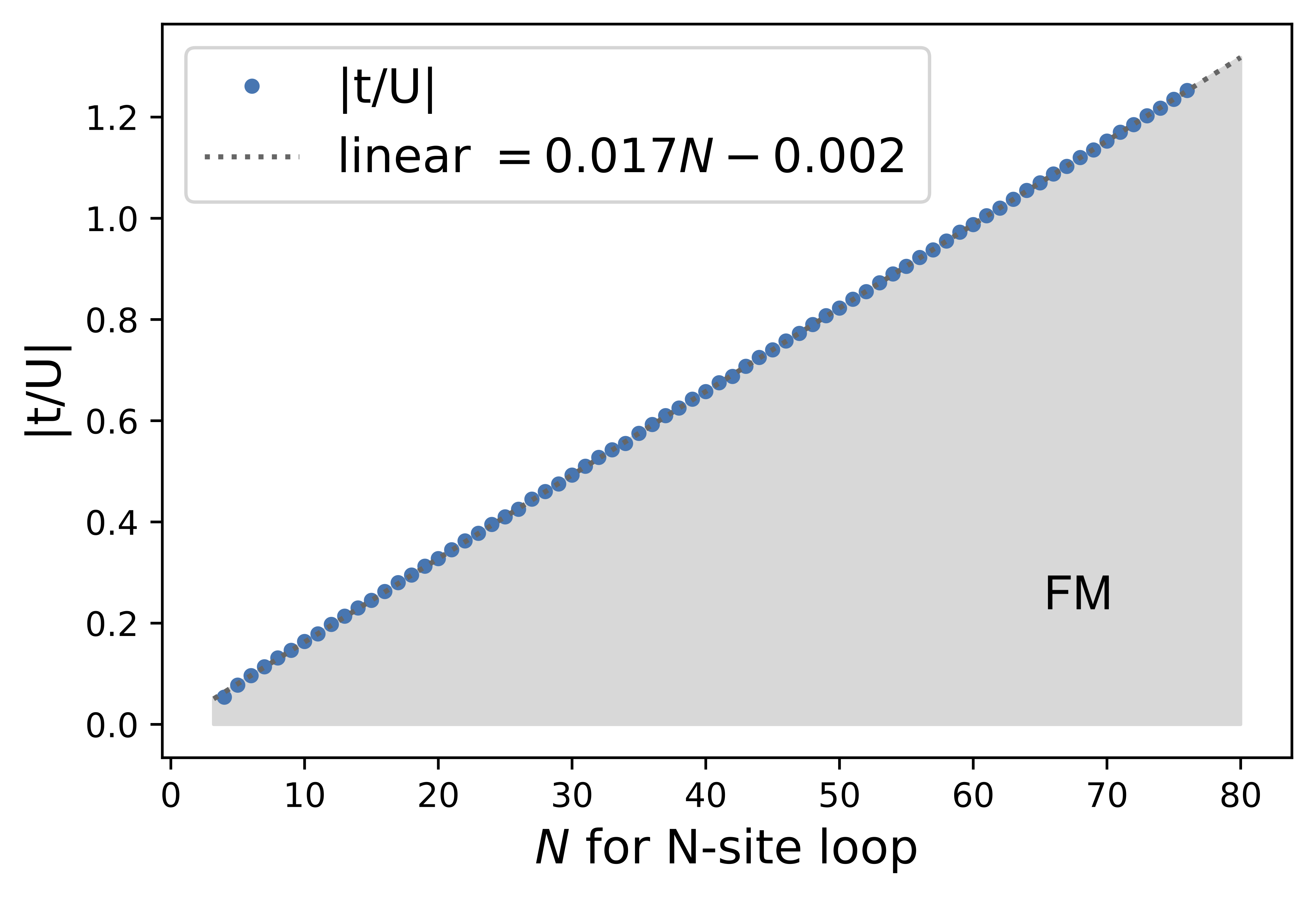}
    \caption{Dependence of the transition point $|t/U|$ on loop size $N_{loop}$ for 3-electron ferromagnetism. As the loop gets larger (while the electron number is fixed at 3) and the $t/U$ transition point to ferromagnetism gets larger. Here only loop sizes accessible with ED are shown. A linear fit is shown.}
    \label{fig:loop_tU}
\end{figure}

When the center site is removed from the $3\times 3$ array, the array becomes an eight-site loop. The many-body wavefunction for this loop is not connected for one-hole in a half-filled band, as shown by Fig. \ref{fig:connected}. As a consequence, NF can't exist. Our calculations confirm this. Interestingly, the 8-site loop does have a ferromagnetic ground state for filling by 3 electrons. For different-size loops ($N_{loop} > 4$), the same, fixed-number-of-electrons (3), ferromagnetic ground state also exists and the transition point increases linearly as the number of sites on the loop increases, shown in Fig. \ref{fig:loop_tU}. In fact, the many-body wavefunction on the loop is connected for a filling of maximum three electrons only. The many-body wavefunction is not connected for any higher electron filling. Thus, a ferromagnetic GS should only be possible for a filling of 3 electrons or less.The one-electron system is trivially polarized. Two-electrons on a loop does not have a transition point to ferromagnetism, so here, we focus on three-electrons loops. This 3-electron FM was also discussed earlier in \cite{Ivan2020} as the only stable kinetic ferromagnetism possible on a loop.

The specific cases for pentagons (five-site loops) and hexagons (six-site loops) were discussed in \cite{Buterakos2023}, which described ferromagnetism appearing in small pentagonal and hexagonal plaquettes at filling factors of roughly 3/10 and 1/4. However, this density dependence appears to be an artifact due to the different-size loops. For pentagon (5 sites) and hexagon (6 sites), 3-electron ferromagnetism would give electronic densities at $3/(N_{loop}\times 2) = 3/10$ or $3/12 = 1/4$. In each case, the ferromagnetism occurs for 3 electron filling. This is not a density effect. This filling for ferromagnetism arises because three-electron filling on a loop allows for a connected many-body wavefunction, while all higher electron fillings do not. For more complex structures like two or three adjacent loops, special fillings at a fixed number of electrons also exist, for example, two adjacent pentagons or hexagons each have 5-electron ferromagnetism, which can be extended to larger loops, like two adjacent nonagons or even two adjacent loops with different sizes. For more discussion, see Appendix \ref{adj_loop}.

The Nagaoka ferromagnetism of $N\times N$ arrays is very different from the GS ferromagnetism of three-electron $N$-site loops. Nagaoka ferromagnetism in an $N\times N$ array occurs for one-hole in a half-filled band. The number of electrons in the NF state increases as $N$ increases. However, for any $N$, the NF state has one hole. In contrast, loop ferromagnetism occurs only for 3 electrons, regardless of the number of loop sites $N$. The transition $t/U$ decreases as an inverse power law of $N$ in $N\times N$ arrays, requiring larger $U$ to achieve ferromagnetism in a larger array. For a loop, the transition point increases linearly with increasing loop size. 

These very different scalings are determined in part by the very different band fillings: 3 electron filling for loops and one-hole in a half-filled band for NF. The loop ferromagnetism occurs for a system with a few (i.e. three) electrons in an empty band. NF occurs for a few holes (i.e. one) in a half-full band. Explaining the different scalings for loop ferromagnetism and NF starts with this difference in band filling.  

\subsubsection{Scaling Arguments for the Transition to Ferromagnetism}
In the following, we will lay out the basic arguments for the two scaling relations to show why they are so different. Additional details and comments are provided in Appendix \ref{scaling}. The linear dependence for loops is explained based on how the hopping energy and the repulsive energy of interaction scale with loop length. For three-electron filling, only states near the bottom of the band are occupied unless $N$ is small. Due to the confinement to the loop, single-particle states near the bottom of the band have energy shifts above the band edge that vary with loop size as $1/N^2$. As a consequence, the kinetic energy difference between the ferromagnetic state and the non-ferromagnetic state should vary with increasing loop length as $t/N^2$. The kinetic energy of the ferromagnetic state, with all three electrons with the same spin will be higher than the kinetic energy of the non-ferromagnetic state with both majority and minority spin electrons. 

The crossover from ferromagnetic to not ferromagnetic occurs when the kinetic energy cost to be ferromagnetic is compensated by the reduction in the repulsive interaction energy when the spins are aligned. There is no interaction energy in the ferromagnetic state because all electrons have the same spin. In the non-ferromagnetic state, there is an interaction energy $U$ for each pair of opposite spin electrons on the same site. The probability that two electrons of opposite spin are on the same site in the loop varies with loop length as $1/N$ and, as a result, the interaction energy varies as $U/N$. The $t/U$ for the transition between the ferromagnetic and non-ferromagnetic states occurs when the kinetic energy cost that varies as $t/N^2$ is compensated by the repulsive interaction energy that varies as $U/N$. The $t/U$ at the transition should scale as $N$ as the loop size varies. This linear scaling for loop ferromagnetism is determined by the three electron filling and the scaling of the kinetic and interaction energies. 

The arguments to support the $1/N^{4}$ size-scaling of the transition $t/U$ for NF in $N \times N$ arrays with one hole in half-filled band are more complicated. Here we summarize the key results. The details are given in Appendix \ref{scaling}. The arguments are easiest to develop in the infinite $U/t$ limit. The transition is identified by why and when the excited, non-ferromagnetic states cross the ferromagnetic ground states as $U$ decreases. In the infinite $U$ limit, the energy levels of states with one hole in a half-filled band can be determined by lowest order degenerate perturbation theory where $|t/U|$ is the small perturbation parameter. In the infinite $U$ limit, the fully polarized, single-hole state is the ground state with the energy for one hole hopping freely in the array. Low-lying partially polarized excited states are shifted up from the ground state by energies that scale between $|t|/N^3$ and $|t|/N^4$ due to how the hole is localized by scattering from minority spins in the array. The fully polarized ground state in the infinite $U$ limit has no repulsive interaction energy so its energy remains fixed as $U$ becomes finite. However, the higher lying, partially polarized states gain exchange energy $-t^2/U$ that lowers their energy, because hopping to double occupied sites becomes possible for finite $U$. The transition to a non-ferromagnetic state occurs when the energy gain due to exchange lowers the excited state energies enough to cross the fully polarized level. This occurs at a $|t|/U$ that scales between $|t|/N^3$ and $|t|/N^4$. This is consistent with the $1/N^{4}$ scaling obtained from the computational results.

\subsubsection{Stark Contrast between Loop and Nagaoka Ferromagnetism}
The scaling arguments point to the very different nature of loop and Nagaoka ferromagnetism. Linear scaling of the transition $|t/U|$ for loop ferromagnetism occurs when the kinetic energy cost for being ferromagnetic is compensated by the reduced repulsive interaction when ferromagnetic. For Nagaoka ferromagnetism in $N\times N$ arrays, the ferromagnetic phase has lower hopping energy making it the ground state in the infinite $U/t$ limit. The not-ferromagnetic phases are lowered in energy due to exchange interactions that couple the not-ferromagnetic states to states that have doubly-occupied sites. This contrasting behavior is connected to loop ferromagnetism occurring at low electron filling and Nagaoka ferromagnetism occurring at single-hole doping.

The $t/U$ transition ratio for NF in $2\times 2$ arrays is $0.053$ and is smaller for larger arrays. The hopping $t$ is much smaller than $U$ at the transition point for all square arrays. The transition point for itinerant ferromagnetism in loops quickly becomes of order 1 with $t$ similar to $U$, as the loop size increases, for all but the smallest loops. This provides another stark contrast between loop and Nagaoka ferromagnetism.

There is one case where loop and Nagaoka ferromagnetism may be the same. The $2\times 2$ array can also be thought as a 4-site loop. In this case, one hole in a half-filled band is three electron filling, so the filling for ferromagnetism is the same whether the array is considered a $2\times 2$ array or a 4-site loop. There is ambiguity in whether ferromagnetism should be consider Nagaoka ferromagnetism or three-electron loop ferromagnetism. For larger systems this distinction becomes clear.  

\subsubsection{Particle Correlation in Loop Ferromagnetism}
\begin{figure}[h!]
    \includegraphics[scale=0.028]{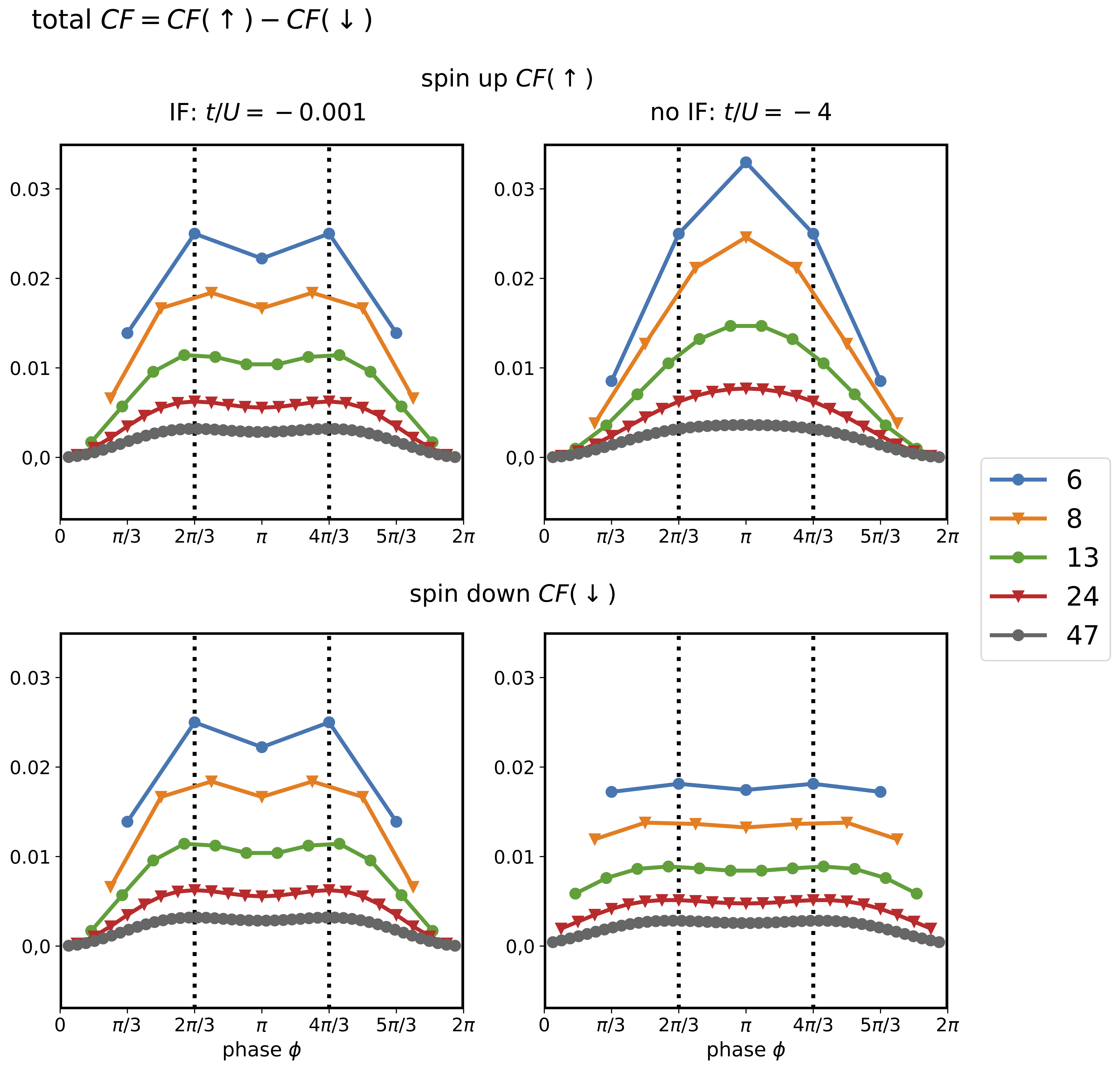}
    \caption{Dependence of the spin-up and spin-down correlation functions versus angle for a loop with three electrons, $s_z = 1/2$ and the fixed electron-spin at $\phi = 0$. Different loop lengths are shown. The left column is for a GS below the $t/U$ transition that is IF. The right column is for a GS above the transition where there is no IF.}
    \label{fig:loop_cf}
\end{figure}

Loop ferromagnetism can be further characterized by evaluating the two-particle correlations. When the loop sites are arranged evenly on a circle, each site position can be parameterized by an angle $\phi$ with $0\leq \phi \leq 2\pi$. Fixing one site on the loop at $\phi = 0$, the correlation functions (CF) for the remaining electrons can be plotted against the angle. In Fig. \ref{fig:loop_cf}, the CFs for a fixed spin up $CF(\uparrow)$ and spin down $CF(\downarrow)$ for three electrons on loops of various sizes are plotted against angle in the left column for a ferromagnetic ground state and in the right column for a nonferrromagnetic ground state. Here we display a GS when $s_z = 1/2$. When the system is in the IF regime, shown here for a small $t/U$ ratio, the spin up/down CFs have peaks at angles $\phi = 2\pi/3$ and $4\pi/3$, with the electrons as far away from each other as possible, and with a total spin of the ground state that is maximum at $S = 3/2$. $CF(\uparrow)$ and $CF(\downarrow)$ are the same, showing that the two other electrons, regardless of spin, are correlated the same way to the fixed spin, as would be expected when interaction is dominant and the wavefunction is connected. When the $t/U$ ratio is large, the saturated ferromagnetism is lost, the two-peak structure disappears, the electrons behave qualitatively differently and the GS is no longer the maximum total spin. For a fixed spin-up electron, the other two electrons are spin up and spin down. With weak correlation, the other spin-up electron stays as far away as possible from the fixed spin-up electron, due to Pauli blockade, while the weakly correlated spin-down electron is more evenly spread around the loop. In contrast, when the spin-down electron is fixed, the remaining two electrons are spread almost uniformly around the entire loop with little correlation as expected when hopping is dominant.

\subsubsection{The Sign of t in Loop Ferromagnetism}
The sign of $t$ matters for loops. For odd-numbered loops, the ferromagnetism for three electrons only occurs when $t<0$ and for two electrons only for $t>0$. For even numbered loops, the ferromagnetism for three electrons exists regardless of the sign of $t$, and there is no ferromagnetism for any other filling. The even-numbered loops are bipartite and a transformation between $t$ and $-t$ can be done. This can also be explained with the exchange interaction coming from itinerant holes, when the exchange interaction is negative, the ground state is ferromagnetic. The specific combination of the sign of $t$, even/oddness of $N$ and electron filling determines the sign of the exchange interaction, and hence determines whether ferromagnetism exists.

\section{\label{summary}Summary and Discussion}  

In this work we have presented theoretical simulations done to probe Nagaoka ferromagnetism in arrays with different geometries. Starting from the $2\times 2$ plaquette, which was experimentally realized in \cite{dehollain2020}, we have shown that perfect $3\times 3$ and $N\times N$ arrays up to $8\times 8$ exhibit NF  for one-hole in a half-filled band, with a $t/U$ ratio for the onset of NF that decreases as the array size N increases, with the transition scaling as $1/n^{4}$. So far, 2D dopant arrays have only been studied for the $3\times 3$ configuration. However, recent advances in fabrication of dopant structures give significant promise for making highly engineered dopant arrays that will be an excellent testbed as an analog quantum simulator for this quantum magnetism .\cite{Wyrick} Extending coupled quantum dot systems to large, complex and extended arrays shows similar promise.

The existence of Nagaoka ferromagnetism or other forms of itinerant ferromagnetism in small arrays crucially depends on the array geometry. One of the geometrical, necessary criteria is connectivity, not only in physical space, but also in the many-body spin-configuration space, which requires that any configuration of spins in the array can be transformed to any other configuration by a series of nearest neighbor hoppings without double occupancy. In $N\times N$ arrays, at least one hole is needed in a half-filled band for wavefunction connectivity at that filling. The $N\times N$ array has saturated ferromagnetism at this filling. A chain of $2\times 2$ square plaquettes with adjacent plaquettes connected at a common corner needs at least two holes in the half-filled band to support wavefunction connectivity and supports saturated ferromagnetism for small $t/U$. In contrast, $N$-site loops have wavefunction connectivity only for three-electron filling and saturated ferromagnetism for three-electron filling.   

The robustness of the saturated ferromagnetism to disorder in $N\times N$ arrays has been tested by considering $3\times 3$ arrays with one site removed. When a corner site is removed, saturated ferromagnetism is still possible for one hole in a half-filled band (which now has one less electron). When one of the edge sites is removed from the $3\times 3$ array, saturated ferromagnetism is no longer possible because wavefunction connectivity is broken. When the center site is removed, the $3\times 3$ array becomes a loop. Saturated ferromagnetism is possible but only for three-electron filling. Higher electron fillings don't have wavefunction connectivity. The transition between different realizations of ferromagnetism when a site is removed can be mapped out by varying the on-site energy of the site to be removed. The phase diagrams associated with these transitions will be discussed in a future publication.

Electron filling plays an essential role in determining the type of ferromagnetism that can occur and the scaling of the transition $|t/U|$ that marks the crossover to ferromagnetism. Loop ferromagnetism occurs for the same, fixed, small electron filling for all loop sizes $N$. The ferromagnetic state occurs for three electron filling but not for more electrons. Nagaoka ferromagnetism in $N\times N$ occurs when there is one hole in a half-filled band. As one consequence the scaling of the transition $|t/U|$ is linear in $N$ for loops and $1/N^{4}$ in arrays. Loop ferromagnetism occurs for a broader range of $U$ as $N$ increases, while for NF, the ferromagnetic state occurs for a smaller range of $U$ as $N$ increases. Scaling arguments support and elucidate the scaling obtained from the computational studies. Three-electron loop ferromagnetism occurs when the kinetic energy costs of becoming ferromagnetic is compensated by the elimination of repulsive interaction in the ferromagnetic state. For NF, the transition out of the ferromagnetic state occurs when the hopping energy cost of the hole, partially localized by scattering from the minority spins is compensated by the exchange energy that is gained when hopping to create doubly occupied states is allowed at finite $U$. 

Interestingly, the first experiment to demonstrate Nagaoka ferromagnetism was for a $2\times 2$ plaquette. For $2\times 2$ plaquette, one hole in a half-filled band requires three electrons. However the $2\times 2$ array is also a loop that displays the three-electron ferromagnetism expected for loops. Ferromagnetism in $2\times 2$ plaquettes could be described as Nagaoka ferromagnetism or as three-electron loop ferromagnetism. These are very different, but it is not clear if one is a better characterization of the ferromagnetism of a $2\times 2$ plaquette.

Our results show that many forms of ferromagnetism should be possible in arrays with various geometries. Both small-scale and extended dopant  and quantum dot arrays should be a rich playground for quantum simulators to explore quantum magnetism.

\begin{acknowledgments}
This material is based upon work supported by the National Science Foundation under Grant No 2240377. 
\end{acknowledgments}

\appendix
\section{Scaling of the transition ratio with system size} \label{scaling}
There are two dramatic differences between Nagaoka ferromagnetism in $N\times N$ arrays and three-electron ferromagnetism in $N$-site loops. First, the band-filling necessary for saturated magnetism is related to the many-body wave function connectivity, which, in turn, is related to the system geometry. The geometries of the $N\times N$ array and the $N$-site are very different, leading to very different criteria for wave function connectivity. In $N\times N$ arrays, at least one hole is needed in a half-filled band to have the wave function connectivity needed for saturated ferromagnetism. For one-hole doping, the electron number scales as $N^2$. Loop ferromagnetism occurs for three-electron filling, independent of loop size. Wave function connectivity in loops is only possible in loops for three-electron filling.

The second essential difference is the scaling of the $t/U$ ratio at the ferromagnetic transition with the system size $N$.
For $N\times N$ arrays, the $t/U$ ratio at the transition scales as $1/N^{3.5}$, as shown in Fig.~\ref{fig:NxN_en}. As the array size $N$ increases, a larger $U$ is needed to transition to ferromagnetism, making it harder to reach ferromagnetism as $N$ increases. For $N$-site loops, Fig.~\ref{fig:loop_tU} shows that the $t/U$ ratio at the transition scales as $N$. As the loop size increases, a smaller $U$ is needed for the transition to ferromagnetism, making it easier to for this transition to take place.

In this appendix we first show how the scaling for three-electron loop ferromagnetism can be understood based on simple arguments about how the difference in system hopping energy before and after the ferromagnetic transition and the corresponding difference in system electron-electron repulsion scale with the system size. Then we show how scaling for NF in an $N\times N$ array can be understood.

\subsection{Scaling of the hopping energy in three-electron loops}

Loop ferromagnetism (LF) occurs for three electron filling.  The three electrons fill states with hopping energies near the band edge unless the loop is very short. Near the band edge, single-particle energies take the form $-a|t|+b|t|/N^2$, where $a$ defines the band edge, $b$ is the coefficient for the quadratic quantum confinement contribution and only the lowest order contribution in $1/N$ is kept. The total hopping energy for three electrons is $-3a|t| + b^{\prime}|t|/N^2$. The quantum confinement contribution for three electrons is determined by $b^{\prime} > 0$. In LF the three electrons are fully polarized, have the same spin and must occupy higher energy states than in states with no loop ferromagnetism (nLF). As a consequence, $b^{\prime}_{LF} - b^{\prime}_{nLF} > 0$ and the hopping energy increases when the three-electron ground state is the LF state. 

\subsection{Scaling of the interaction energy in three-electron loops}

In the LF state, the three electrons have the same spin and there is no interaction energy. In contrast, in the nLF state, the state is only partially polarized and the interaction energy arises from the repulsive interaction $U$ between each pair of electrons with opposite spins that are on the same site. The probability that two electrons are on the same site in the loop scales as $1/N$. Thus the repulsive interaction energy scales as $n_{int}U/N$, where $n_{int}$ is the number of interacting pairs and is of order 1. 

\subsection{Scaling the transition from nLF to LF in three-electron loops}

The crossover from nLF to LF occurs when the the additional hopping energy cost compensates the reduction of the interaction energy:
\[(b^{\prime}_{LF} - b^{\prime}_{nLF})|t|/N^2 = n_{int}U/N \]
As a result of the scaling of the hopping and interaction energies, $|t|/U$ at the phase transition scales as $N$. This is the scaling obtained from diagonalization of the three-electron Hamiltonian for loops with different sizes $N$.

\subsection{Using the states of the $N\times N$ array in the large $U$ limit to establish the scaling of the transition}

\begin{figure}[h!]
    \includegraphics[scale=0.45]{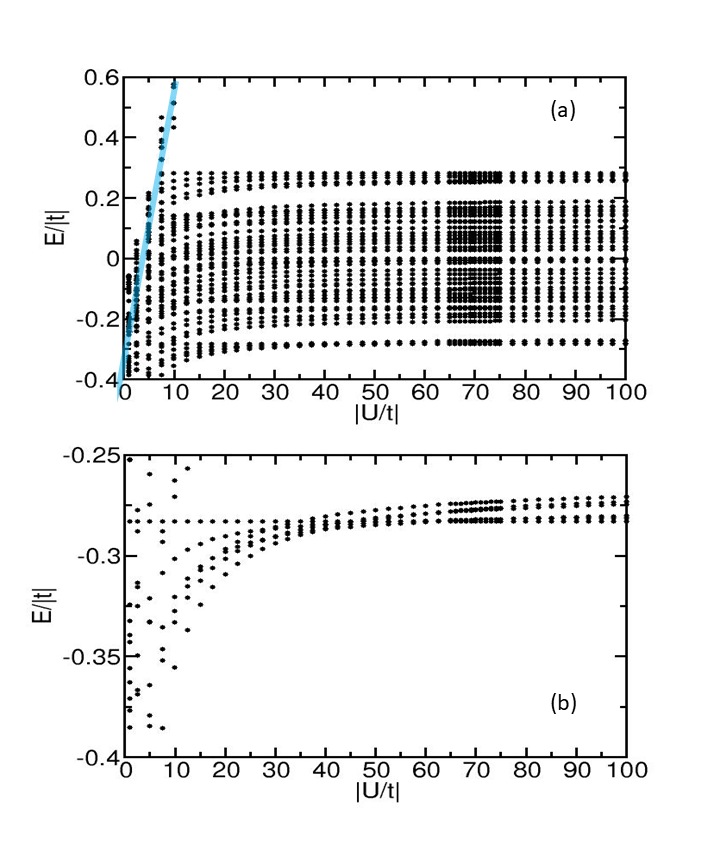}
    \caption{Energy levels of a $3\times 3$ array with one hole in a half-filled band with seven majority spins and one minority spin. (a) Full spectrum, dependence on $|U/t|$. The region with a high-density of points is the region where the crossover between NF and nNF occurs. The blue-shaded region highlights the energies of states with one doubly-occupied site. (b) Blow up of the region near the transition.}
    \label{fig:appendix}
\end{figure}

To establish the scaling of the transition point in $N\times N$ arrays, it is easiest to start in the infinite $U$ limit and then determine how the energy levels shift as $U$ becomes finite and moves to the transition point. The dependence on $|U/t|$ of the energy-level structure for a $3\times 3$ array with one hole in a half-filled band is shown in Fig. \ref{fig:appendix} with the transition from no Nagaoka ferromagnetism (nNF) to Nagaoka ferromagnetism (NF) occurring in the region with the higher density of points. Energy levels are shown for the spin configuration with seven majority spins and one minority spin. The level structure for other spin configurations is similar, except with more levels. The lowest energy levels in the infinite $U$ limit are independent of $U$ because these are states with no doubly occupied sites. States with one doubly occupied site are highlighted in blue and can be identified as the lowest energy states with energies that increase linearly with $|U/t|$. Higher energy states with more than one doubly occupied site are not shown. The level crossing that determines the nNF to NF transition occurs in the levels for states with no doubly occupied sites. 

The level structure is easiest to understand in the large $U$ limit with a simple structure above and near the crossover. The levels are easy to follow as $U$ becomes finite and moves toward the crossover regime. Below the crossover regime, the levels exhibit many crossings making it hard to follow the levels as $U$ varies. To understand the scaling of the transition $|U/t|$ on $N$, we will follow the levels from the infinite $U$ limit down to the crossover regime. 

In the large $|U/t|$ limit, it is more instructive to work with the Hamiltonian $\mathcal{H}$ scaled by $U$:

\begin{equation}
\mathcal{H}_{sc} = \mathcal{H}/U = (t/U)\sum_{<i,j>, \sigma} c^{\dag}_{i,\sigma}c_{j,\sigma} + \sum_i n_{i,\uparrow} n_{i,\downarrow .}
\end{equation}The first term is the hopping term and is a small contribution that can be treated as a perturbation. The second term, independent of $U$, is the large contribution that determines the form of the states in the large $|U/t|$ limit. In the infinite $U$ limit, all states with $n_{do}$ doubly occupied sites have an energy of $n_{do}$. The states with no doubly occupied sites form a degenerate ground state subspace. This degeneracy can be broken by using degenerate zero-order perturbation theory and diagonalizing the hopping term in the subspace of states with no double occupancy.

Nagaoka ferromagnetism occurs for one hole in a half-filled band. When all spins are polarized in the same direction, one-hole doping corresponds to one hole in the completely filled band of majority-spin electrons. In the following we will describe states by the number of holes in the filled majority-spin band plus the number of minority-spin electrons. For simplicity we describe such a state by the number of holes and electrons. To have the band filling needed for NF, we must have one more hole than than the number of electrons. In the subspace where there are no doubly occupied sites, each electron must sit on the same site with one of the holes. There will be a bound electron-hole pair for each electron and one extra free hole. An electron can't move away from the hole that it is paired with unless it can hop directly to the site occupied by the free hole. The holes bound to electrons can't move away from the electron they are paired with because that would create a doubly occupied site. Only the free hole can move by hopping.

\subsection{Scaling of the hopping energy for $N\times N$ arrays in the infinite $U$ limit}

Using the model just described, we now analyze the energy-level structure of states with band filling needed for NF in the infinite $U$ limit. The lowest energy state in the subspace with $n_{do} = 0$, should be the fully polarized state with one hole and no electrons. In this state, the hole moves freely in the $N\times N$ array with the motion determined by the hopping term in $\mathcal{H}_{sc}$. The energy levels are those of a free particle hopping in an $N\times N$ array. The lowest level at the band edge has an energy that scales with $-|t|/U$ (as $-|t|$ for the Hamiltonian without scaling). The excited levels close to the band edge are shifted to higher energy as $|t|/UN^2$ by quantum confinement effects and levels further away are split by $|t|/UN$ due to the density of levels. As a consequence, the excited levels close to the band edge should have a splitting from the band edge that scales between $|t|/UN$ and $|t|/UN^2$ (between $|t|/N$ and $|t|/N^2$ for the Hamiltonian without scaling).

The simplest states with partial polarization would have two holes and one electron, with one of the holes bound to the electron on the same site. The bound pair can't move, unless it exchanges with the free hole, which is only possible when the hole and bound pair are nearest neighbors. The unbound hole is nearly free, able to move through the lattice except where the bound pair is. The bound pair acts to partially localize the nearly free hole, increasing the energy of the nearly free hole. As a consequence the fully polarized ground state, with no bound pairs, is the lowest energy state in the infinite $U$ limit. There are on the order of $N^2$ unique positions where the bound pair can sit. The existing single-hole energy levels should be further split into different levels for each partially localized state with level splittings scaling as $|t|/N^4$ near the band edge and $|t|/N^3$ away from the band edge. As a consequence, the excited levels close to the band edge should have a splitting from the band edge that scales between $|t|/UN^3$ and $|t|/UN^4$ (between $|t|/N^3$ and $|t|/N^4$ for the Hamiltonian without scaling).

More complicated states with less polarization would have one nearly free hole and multiple bound electron-hole pairs. The same arguments about the bound pairs localizing the nearly-free hole should apply. As a consequence, the excited levels close to the band edge should have a splitting from the band edge that scales between $|t|/UN^3$ and $|t|/UN^4$ (between $|t|/N^3$ and $|t|/N^4$ for the Hamiltonian without scaling). In addition, the same arguments imply that the fully polarized ground state has lower energy than any partially polarized state in the infinite $U$ limit. 

\subsection{Scaling the exchange interaction for $N\times N$ arrays for finite repulsive interaction}

These arguments define the level structure and the scaling of the level splitting between the fully polarized ground state and the excited states with no doubly occupied sites in the infinite $U$ limit. We now need to determine how this leveling splitting changes as $U$ becomes finite. To do this we need to determine the second-order change in energy of the excited states due to hopping term in $\mathcal{H}_{sc}$. The energy of the fully polarized ground state (for large $U$) is independent of $U$ because there in no same-site occupation and no contribution from $U$ in this case. For excited states the second-order change arises from hopping to a single doubly-occupied site. For $\mathcal{H}_{sc}$, this second-order energy shift scales as $-(t/U)^2$ (for $\mathcal{H}$, this second-order energy shift scales as $-t^2/U$. This second order change due to exchange coupling to the states with one doubly-occupied site is revealed in Fig.\ref{fig:appendix} as level repulsion between states with one doubly-occupied site and the low-energy excited states with no doubly-occupied sites.   

\subsection{Scaling the transition from nNF to NF for $N\times N$ arrays}

The transition from NF to nNF as $U$ decreases from the infinite $U$ limit occurs when the energy shift from the exchange coupling to the states with a doubly-occupied site closes the gap between the excited state and the fully polarized ground state. Using $\mathcal{H}_{sc}$, the second-order exchange $-(t/U)^2$ closes the gap which scales between $|t|/UN^3$ and $|t|/UN^4$. As a consequence, the scaling of $|t|/U$ at the transition should be between $1/N^4$ and $1/N^3$. This is consistent with our calculations for different $N$ which show a $1/N^{4}$ scaling. When we include all values of $N$ to determine the scaling, the scaling is closer to $1/N^{3.5}$ scaling. The scaling for the larger $N$ should depend of the state splitting near the band edge and should have a scaling closer to $1/N^4$, consistent with the scaling we get for larger $N$ in the computational results. When the smaller $N$ are included in the scaling, then the small $N$ results will involve states further from the band edge which provide a $1/N^3$ splitting and scaling. The scaling using all $N$ should lie between $1/N^3$ and $1/N^4$, consistent with the $1/N^{3.5}$ scaling seen in the calculated results when all $N$ are considered.

\section{Other Ferromagnetism} \label{other_ferromagentism}

\subsection{NF in Non-squares Arrays}\label{other-NF}
We find that NF exists in rectangular and hexagonal arrays, when the squares or hexagons are connected by sharing a side. See the following Table \ref{tab:other-NF}.
\begin{table}[h!] 
    \centering
    \begin{ruledtabular}
    \begin{tabular}{p{1.5cm}p{1.3cm}p{0.9cm}p{0.7cm}p{1.5cm}p{1.5cm}}
        Geometry & Sketch & NF/IF & $\# e$& Transition & Details\\ \hline
        \makecell{2x3 \\ (2 side-sharing sq)} & \includegraphics[height=0.4in]{2by3.png} & NF  & 5 & $ 0.019$ & 1-hole NF \\ \hline
        \makecell{2x4 \\ (3 side-sharing sq)} & \includegraphics[height=0.4in]{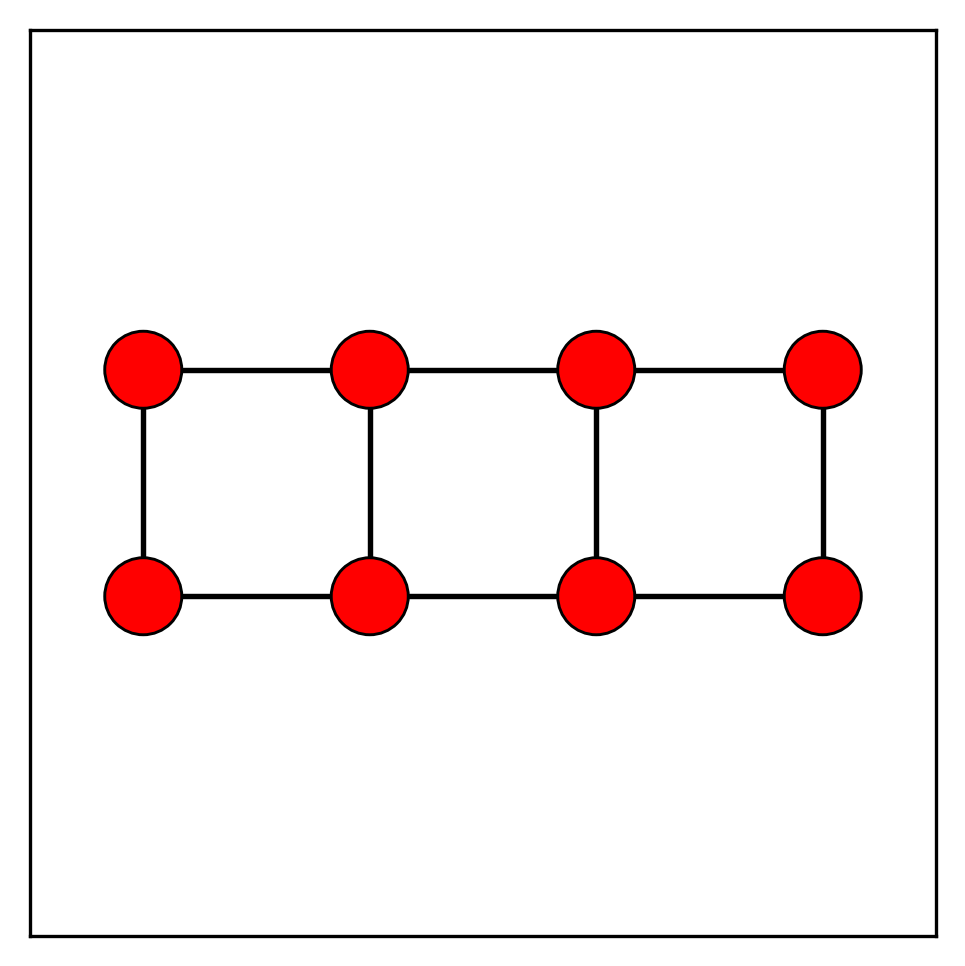} & NF  & 7 & $ 0.012$ & 1-hole NF \\ \hline
        \makecell{2x5 \\ (4 side-sharing sq)} & \includegraphics[height=0.4in]{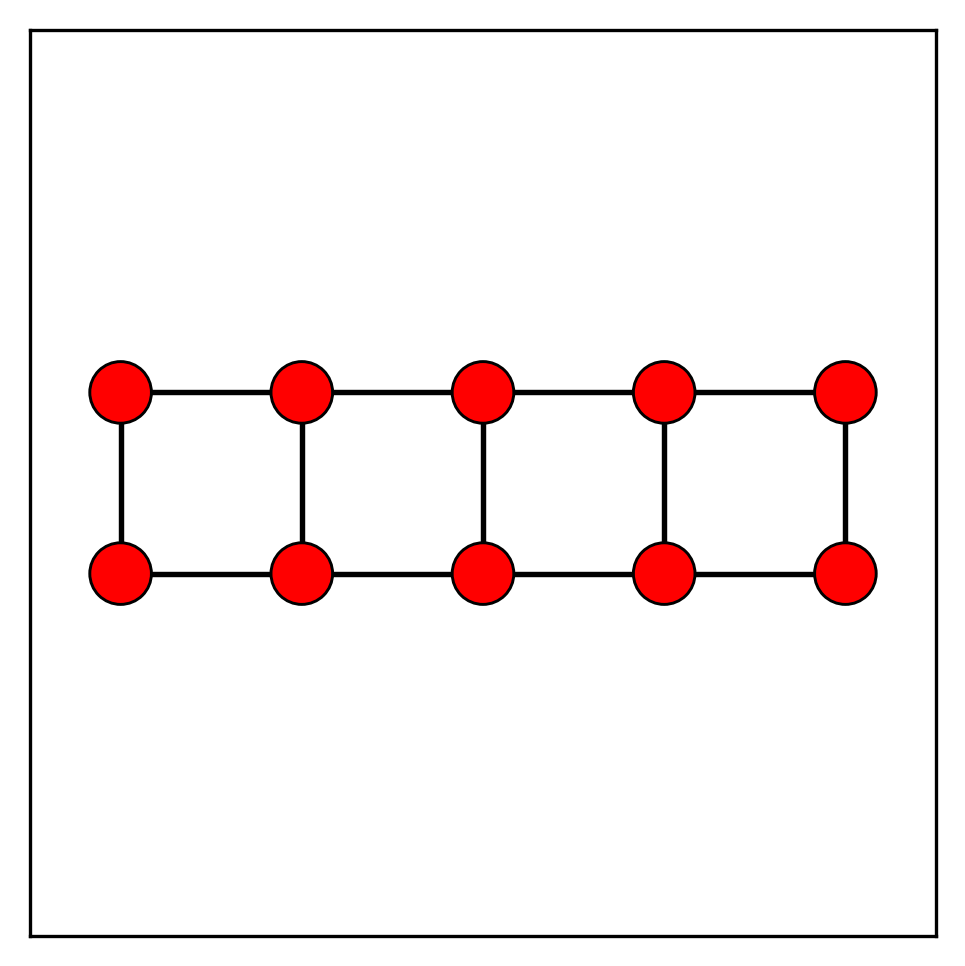} & NF  & 9 & $0.008$ & 1-hole NF \\ \hline
        \makecell{2 side-sharing \\ hexagon} & \includegraphics[height=0.4in]{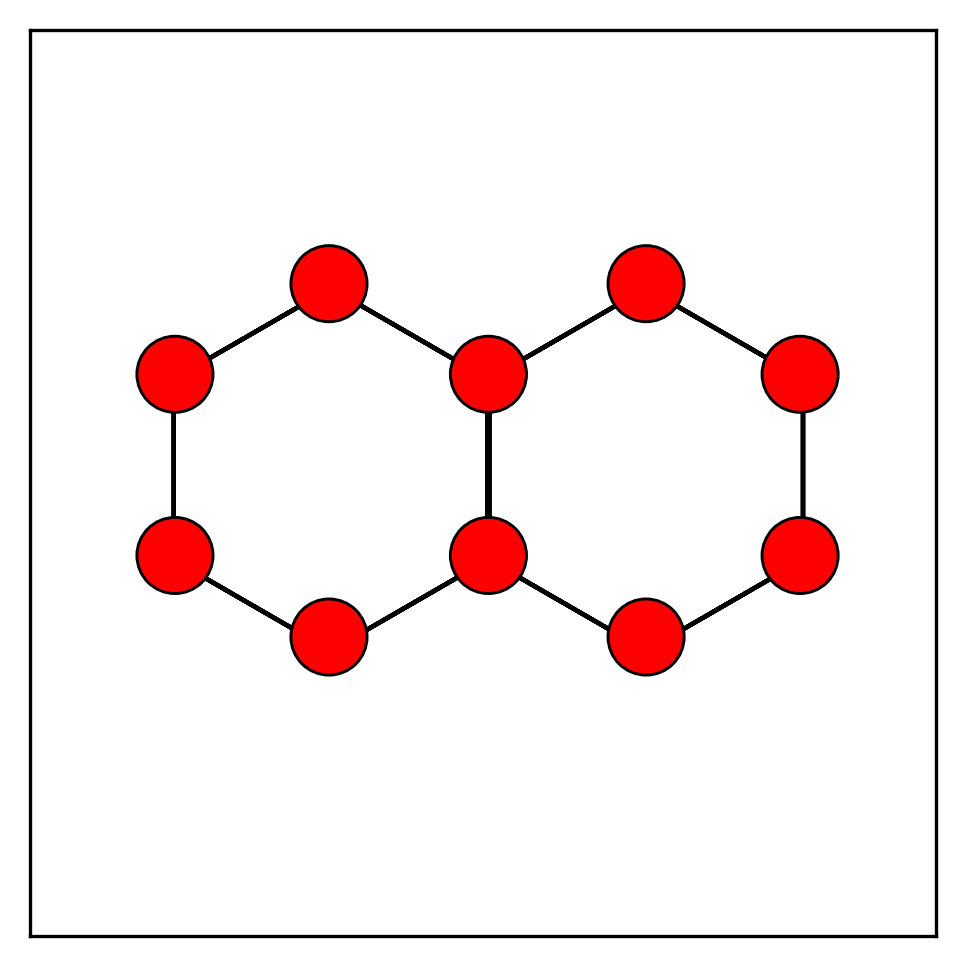} & \makecell{NF \\ IF} & \makecell{9 \\ 5}  & \makecell{$0.0022$ \\ $0.0405$} & \makecell{1-hole NF \\ 5e-IF} \\ \hline
        
    \end{tabular}
    \caption{Summary table for other geometries with NF, their sketch, number of electrons, and $t/U$ transition point.}
    \label{tab:other-NF}
    \end{ruledtabular}
\end{table}

For hexagonal arrays, the one hole ferromagnetism is NF, while the 5 electron itinerant ferromagnetism comes when the structure is two loops sharing a side. This 5-electron ferromagnetism can be extended to other numbers of sites on the loop, we are able to calculate up to two 18-site loops sharing a side. 

\subsection{Two-hole IF in plaquette sharing corners} \label{2-hole}
The two-hole IF extends to larger chains of squares, see following Table \ref{tab:two-hole}.

\begin{table}[h!]
    \centering
    \begin{ruledtabular}
    \begin{tabular}{p{1.5cm}p{1.3cm}p{0.9cm}p{0.7cm}p{1.5cm}p{1.5cm}}
        Geometry & Sketch & NF/IF & $\# e$& Transition & Details\\ \hline
        \makecell{no opposite corner \\ (2 corner-sharing sq)} & \includegraphics[height=0.4in]{noOpC.png} & IF  & 5 & $0.0127$ & 2-hole IF \\ \hline
        \makecell{3 corner-sharing sq} & \includegraphics[height=0.4in]{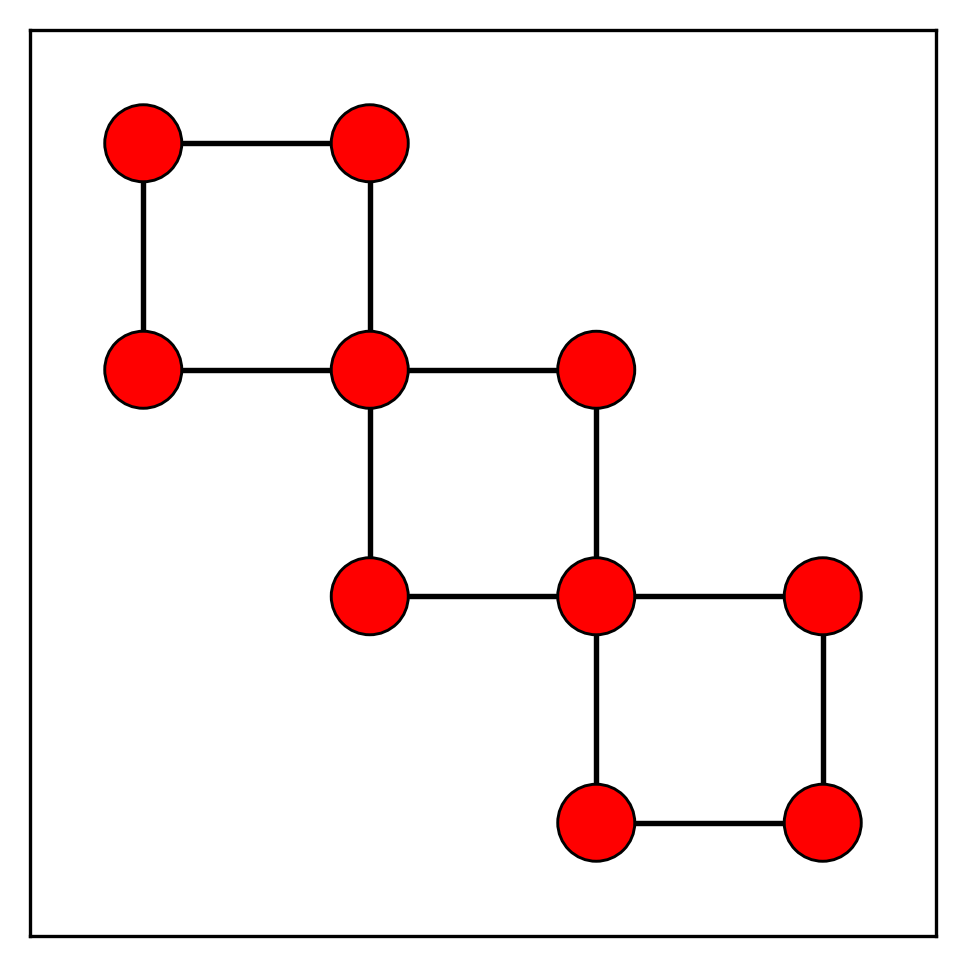} & IF  & 8 & $0.0015$ & 2-hole IF \\ \hline
        \makecell{4 corner-sharing sq} & \includegraphics[height=0.4in]{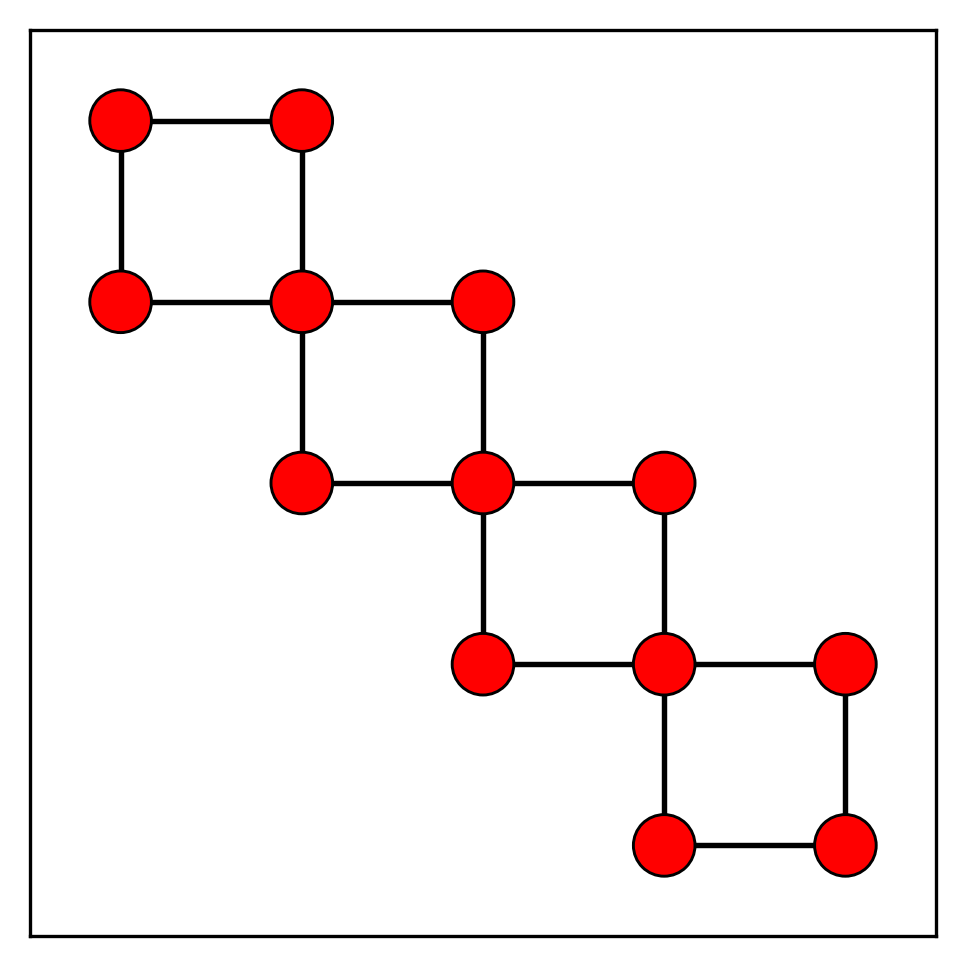} & IF  & 11 & $0.0011 $ & 2-hole IF \\ \hline
        \makecell{5 corner-sharing sq} & \includegraphics[height=0.4in]{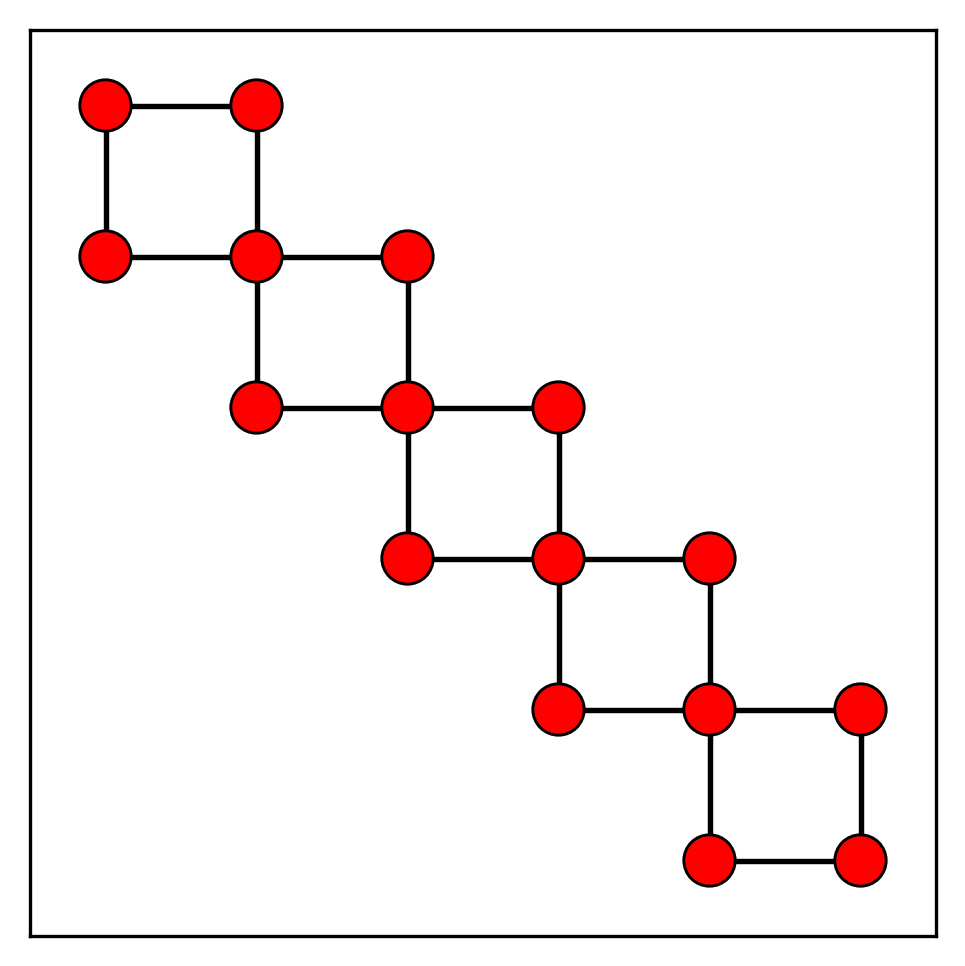} & IF  & 14 & $0.00025$ & 2-hole IF \\ \hline
    \end{tabular}
    \caption{Summary table for two-hole IF, their sketch, number of electrons, and transition point if applicable.}
    \label{tab:two-hole}
    \end{ruledtabular}
\end{table}
These chains of plaquettes sharing corners need not be in a straight chain with the connecting corners along a common diagonal as shown in Table \ref{tab:two-hole}. For example, the connecting corners could be along a line made from a side of each plaquette. More complicated geometries for chains of squares also show IF if certain symmetry requirements are satisfied. This will be discussed in a future publication.  

\subsection{Adjacent Loops} \label{adj_loop}
For small arrays made of adjacent loops, there are special fillings which give rise to saturated ferromagnetism. The fillings are related to the number of loops present in the system. For two adjacent loops (calculated by ED for up to two 14-site loops), with the two loops sharing one hopping and two sites, there exists 5-electron ferromagnetism when the two loops are about the same size. The dependence of the transition point on total system size N is shown below. Similar situation also happens for three adjacent loops with 7-electron ferromagnetism for certain geometries. How the loop number and special fillings are related will require further investigation. 
\begin{figure}[h!]
    \centering
    \includegraphics[scale=0.05]{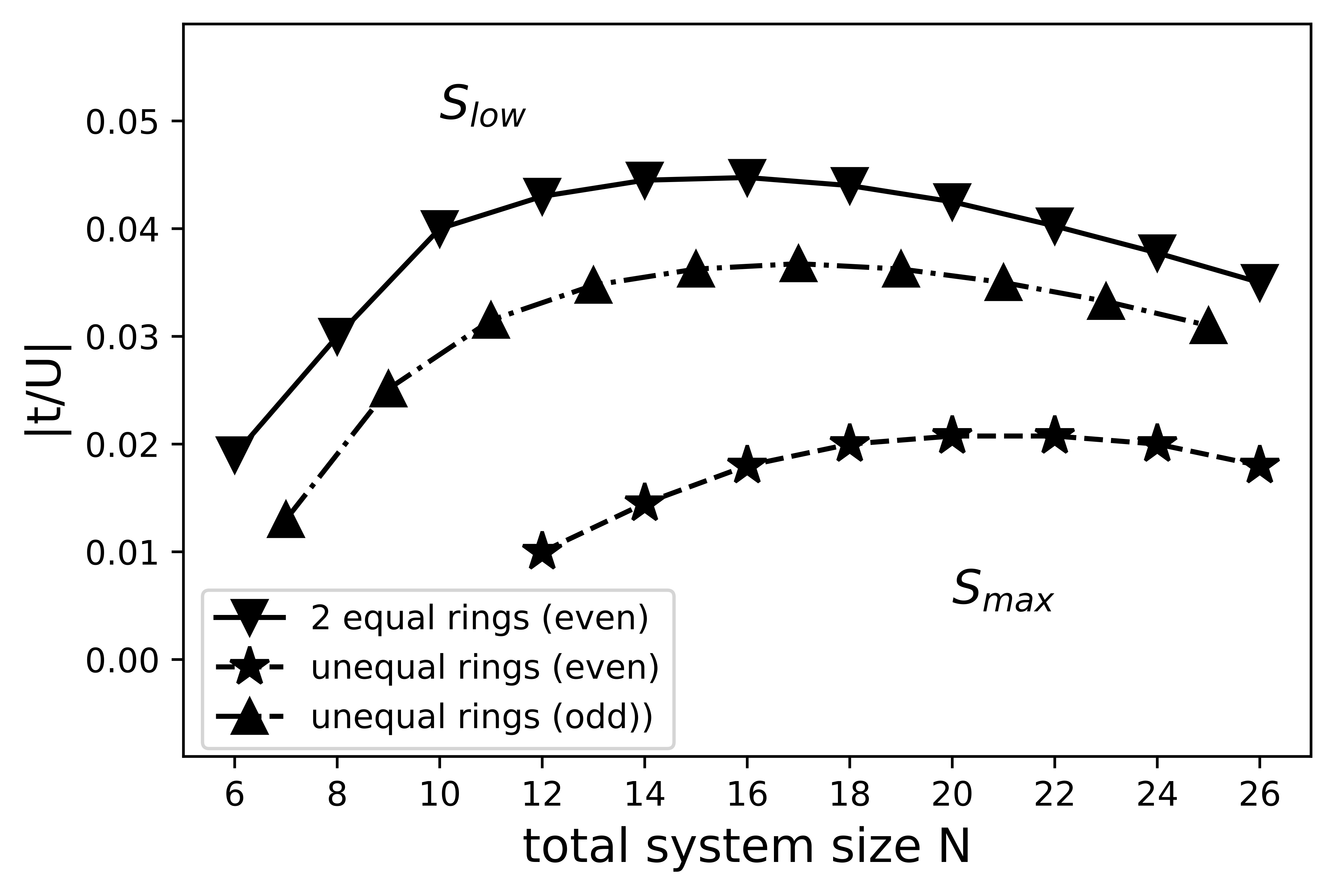}
    \caption{Two adjacent loops. Transition point $|t/U|$ versus total system size $N$ }
    \label{fig:2_rings_adj}
\end{figure}

\subsection{Robustness to extended Coulomb interaction} \label{Coulomb}

In actual experiments with dopant arrays \cite{wang2022}, long-range Coulomb interactions ($V$) exist, and NF is robust for nearest neighbor Coulomb interaction $V_{nn}$ up to $V_{nn}/U \approx 0.017$ in $3\times 3$ arrays. This has been checked with calculations not reported here.

\bibliography{apssamp}

\end{document}